\documentclass[usegraphicx]{mn2e}
\usepackage{amsmath}
\usepackage{amssymb}
\usepackage{times}
\usepackage{subfigure}
\usepackage{epsfig}

\renewcommand{\descriptionlabel}[1]%
  {\hspace{\labelsep}\textbf{#1}}

\title[Examining the variable stars in NGC 5024]
     {Exploring the variable stars in the globular cluster NGC~5024 (M53): New RR Lyrae and SX Phoenicis stars\thanks{Based on
  observations collected at the Indian Astrophysical Observatory, Hanle, India.}}

\author[A. Arellano Ferro et al.]
{A. Arellano Ferro$^{1}$\thanks{E-mail:armando@astro.unam.mx},  
R. Figuera Jaimes$^{1}$\thanks{E-mail:rfiguera@astro.unam.mx},
Sunetra Giridhar$^{2}$\thanks{E-mail:giridhar@iia.res.in}, 
D.M. Bramich$^{3}$\thanks{E-mail:dan.bramich@hotmail.co.uk}, 
\and
J.V. Hern\'andez Santisteban$^{1}$\thanks{E-mail:jhernand@astro.unam.mx}, 
K. Kuppuswamy$^{2}$\thanks{E-mail:kuppuswamy@iia.res.in}.
  \medskip
\\$^{1}$Instituto de Astronom\1a, Universidad Nacional Aut\'onoma de M\'exico\\
\\$^{2}$Indian Institute of Astrophysics, Koramangala 560034, Bangalore, India\\
\\$^{3}$European Southern Observatory, Karl-Schwarzschild-Stra$\beta$e 2, 85748
Garching bei M\"{u}nchen, Germany
}

\begin{document}
 
\date{Accepted . Received ; in original form }

\pagerange{\pageref{firstpage}--\pageref{lastpage}} \pubyear{2011}

\maketitle 

\label{firstpage}

\begin{abstract}
We report CCD $V$ and $I$ time series photometry of the globular cluster NGC 5024 (M53).
The technique of difference image analysis has been used which enables photometric precisions
better than 10 mmag for stars brighter than $V \sim 18.5$ mag even in the crowded central 
regions of the cluster. 
The high photometric precision has resulted in the discovery of two new RR1 stars and thirteen
SX Phe stars. A detailed identification chart is given for all the variable stars in the 
field of our images of the cluster.
Periodicities were calculated for all RR Lyraes and SX Phe stars
and a critical comparison is made with previous determinations.

Out of four probable SX Phe variables reported by D\'ek\'any \& Kov\'acs (2009),
the SX Phe nature is confirmed only for V80, V81 is an unlikely case while V82 and V83 remain as dubious cases. Previous misidentifications of three variables are corrected. 
Astrometric positions with an uncertainty of 
$\sim0.3$ arcsec are provided for all variables.

The light curve Fourier decomposition of RR0 and RR1 is discussed, we find a mean metallicity of 
[Fe/H]=$-1.92 \pm 0.06$ in the scale of Zinn \& West from 19 RR0 stars. The true distance moduli  
$16.36\pm 0.05$ and $16.28\pm 0.07$ and the 
corresponding distances $18.7 \pm 0.4$ and $18.0 \pm 0.5$ kpc
are found from the RR0 and RR1 stars respectively. These values are in agreement with the theoretical period-luminosity 
relations for RR Lyrae stars in the $I$ band and with recent luminosity determinations for the 
RR Lyrae stars in the LMC.

The age of 13.25$\pm$0.50 Gyr (Dotter et al. 2010), for NGC~5024, $E(B-V)=0.02$
and the above physical parameters of the
cluster, as indicated from the RR0 stars, produce a good isochrone fitting to the observed CMD.

The PL relation for SX Phe stars in NGC~5024 in the $V$ and $I$ bands is discussed in the light of the 13 
newly found SX Phe stars and their pulsation mode is identified in most cases.

\end{abstract}
      

\section{Introduction}

This paper is part of a series on CCD photometry of globular clusters in which
we employ the technique of difference image analysis (DIA) to produce precise time 
series photometry of individual stars down to $V \sim$ 19.5 mag even in crowded regions. The DIA photometry has 
proved to be a very useful tool in obtaining high 
quality light curves of known variables, and for discovering and classifying new variables. We also estimate the mean [Fe/H], $M_V$ and $T_{\rm eff}$
for the RR Lyrae stars by the technique of light curve Fourier decomposition
(see e.g. Arellano Ferro et al. 2010; Bramich et al. 2011 and references there in). 

The globular cluster NGC 5024 (M53) RA~$\alpha = 13^{\mbox{\scriptsize h}} 12^{\mbox{\scriptsize m}} 55.3^{\mbox{\scriptsize s}}$,
Dec.~$\delta = +18\degr 10\arcmin 09\arcsec$, J2000; $l = 332.97\degr$, $b = +79.77\degr$ 
 is located in the intermediate Galactic halo at
$Z=$ 17.5 kpc, R$_G$=18.3 kpc (Harris 1996) hence its reddening is very low ($E(B-V) \sim 0.02-0.08$). With [Fe/H]~$\sim -2.0$, it is among the most metal deficient globular clusters in the Galaxy. 

The present edition of the Catalogue of Variable stars in Globular Clusters (CVSGC)
(Clement et al. 2001) lists 90 variable stars in NGC~5024, 62 of which are RR Lyrae stars of either 
RR0 or RR1 type, 11 confirmed and 4 suspected SX Phe, 2 confirmed and 6 suspected semi-regular variables at the Red Giant tip, 3 suspected 
slow irregular variables (or Lb's), and 2 stars which were found $a~posteriori$ not to be variables 
(V22 and V39).

The cluster has been the subject of two recent studies based on CCD photometry by
Kopacki (2000) (K00) and D\'ek\'any \& Kov\'acs (2009) (DK09). Our images cover a larger 
field than those of K00 and are a bit smaller
than those of DK09. However the resolution of our images  is much better than those in these previous works: 0.296 arcsec pix$^{-1}$, compared to 0.688 arcsec pix$^{-1}$ for K00 and 1.026 arcsec pix$^{-1}$
for DK09, leading to better deblending of stars in crowded regions, hence the number of measured stars almost triples to about 8910 in $V$ and 8619 in $I$. For these reasons we feel that our study can contribute to a better understanding of the properties of the
many variables in NGC 5024, to their field identification and to the search for new variables.
We also note that although our observations obtained in 2009 and 2010 represent an extension of the observations 
by K00 and DK09 acquired in 1998-1999 and 2007-2008 respectively, we choose not to combine the data for period 
finding purposes since a close inspection of the light curves from K00 showed that they have a large intrinsic 
scatter and some difference in zero point. Likewise the precision of the
light curves measured by DK09 particularly in crowded regions, most likely due to the low resolution of their 
images, was not very good.

The major goal of our studies of variable stars in globular clusters has been
to obtain internally precise light curves of RR Lyrae stars and use them to estimate mean values of [Fe/H] and $M_V$ by their Fourier decomposition into 
their harmonics. Two colour photometry permits to estimate the age of the cluster and to discover 
and discuss the nature of some variables from their position in the CMD.

In this paper we present a complete census of the variables in the cluster, and estimate the mean values of the metallicity and distance
to the cluster using monoperiodic RR Lyrae stars. We
provide astrometric positions of the variable stars with an accuracy of $\sim$ 0.3 arcsec.

In Sect. 2 we describe the observations and data reductions. In Sect. 3 a detailed discussion is presented on the variability and identification of many individual objects and some new variables are presented. In Sect. 4 
we calculate  the metallicity and absolute magnitude of the RR Lyrae stars using the Fourier light curve decomposition method and discuss the transformations 
to homogeneous scales. In Sect. 5 we present the newly found SX Phe stars and calibrate their PL relationship. 
In Sect. 6 brief comments on the long term variables in the cluster are made in the light of our present data. 
Section 7 contains comments on the age of NGC 5024 and in Section 8  our findings are summarized.

\section{Observations and Reductions}
\label{sec:Observations}

\subsection{Observations}

The observations employed in the present work were performed using the Johnson $V$ and $I$ filters on April 17, 18 and 19, 2009, on January 22 and 23,
February 21 and 22, March 7 and 8, April 7 and 25, and on May 6 2010. A total of 177 epochs in the $V$ filter and 184
in the $I$ filter were obtained.
The 2.0m telescope of the Indian Astronomical Observatory (IAO), Hanle, India, located at 4500m above sea level was used. The estimated seeing was $\sim$1 arcsec.
The detector was a 
Thompson CCD of 2048 $\times$ 2048 pixels with a pixel
scale of 0.296 arcsec/pix and a field of view 
of approximately $10.1 \times 10.1$ arcmin$^2$.

\subsection{Difference Image Analysis}

We employed the technique of difference image analysis (DIA) to extract high precision photometry
for all point sources in the images of NGC 5024 (Alard \& Lupton 1998; Alard 2000;
Bramich et al. 2005).
We used a pre-release version of the {\tt DanDIA}\footnote{
{\tt DanDIA} is built from the DanIDL library of IDL routines available at http://www.danidl.co.uk}
pipeline for the data reduction process (Bramich et al., in preparation) which includes a new algorithm
that models the convolution kernel matching the point-spread function (PSF) of a pair of images of the same field as a discrete pixel array
Bramich (2008).

The {\tt DanDIA} pipeline was used to perform a  number of image calibration steps that include the bias and flat corrections, the cosmic ray cleaning of all raw images, and the creation of a reference image for each filter by stacking a set of the best-seeing calibrated images taken on a single night.
In each reference image, we measured the fluxes (referred to as reference fluxes) and positions of all PSF-like objects (stars)
by extracting a spatially variable (with polynomial degree 3) empirical PSF from the image and fitting this PSF to each detected object. The detected stars in each image in the time-series
sequence were matched with those detected in the corresponding reference image, and a linear transformation was derived which was used to register each image with
the reference image. For each filter, a sequence of difference images was created by subtracting the relevant reference image, 
convolved with an appropriate
spatially variable kernel, from each registered image. The differential fluxes for each star detected in 
the reference image were measured on each difference image. 
Light curves for each star were constructed by calculating the total flux $f_{\mbox{\scriptsize tot}}(t)$ 
in ADU/s at each epoch $t$ from:
\begin{equation}
f_{\mbox{\scriptsize tot}}(t) = f_{\mbox{\scriptsize ref}} + \frac{f_{\mbox{\scriptsize diff}}(t)}{p(t)}
\label{eqn:totflux}
\end{equation}
where $f_{\mbox{\scriptsize ref}}$ is the reference flux (ADU/s), $f_{\mbox{\scriptsize diff}}(t)$ is the differential flux (ADU/s) and
$p(t)$ is the photometric scale factor (the integral of the kernel solution- see Section 2.2 of Bramich 2008). Conversion to instrumental magnitudes was achieved using:
\begin{equation}
m_{\mbox{\scriptsize ins}}(t) = 25.0 - 2.5 \log (f_{\mbox{\scriptsize tot}}(t))
\label{eqn:mag}
\end{equation}
where $m_{\mbox{\scriptsize ins}}(t)$ is the instrumental magnitude of the star at time $t$. Uncertainties were propagated in the correct analytical fashion.

The above procedure and its caveats have been described in detail in a recent paper 
(Bramich et al. 2011) and the interested reader is refered to it for the relevant details.

\begin{figure} 
\includegraphics[width=8.cm,height=8.cm]{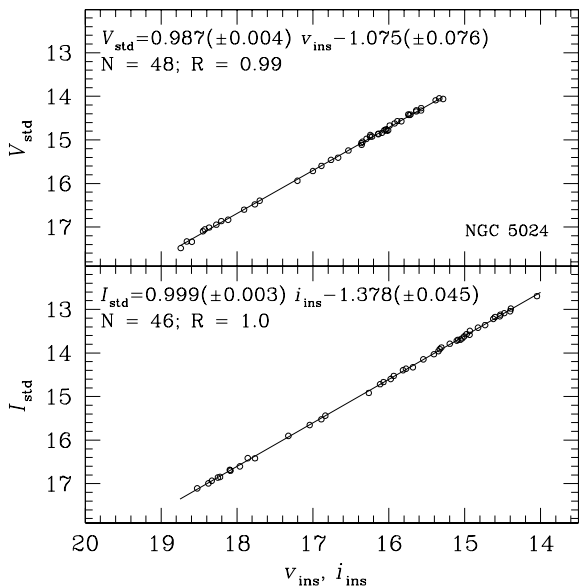}
\caption{Transformation relation between the instrumental $v,i$ and the standard $V,I$ photometry using the standards of Stetson (2000)}
    \label{transVI}
\end{figure}

\begin{table*}
\caption{Time-series $V$ and $I$ photometry for all the confirmed variables in our field of view.
         The standard $M_{\mbox{\scriptsize std}}$ and instrumental $m_{\mbox{\scriptsize ins}}$ magnitudes are listed in columns 4 and 5, respectively, corresponding to the variable
         star, filter, and epoch of mid-exposure listed in columns 1-3, respectively. The uncertainty on $m_{\mbox{\scriptsize ins}}$ is listed
         in column 6, which also corresponds to the uncertainty on $M_{\mbox{\scriptsize std}}$. For completeness, we also list the quantities $f_{\mbox{\scriptsize ref}}$, $f_{\mbox{\scriptsize diff}}$
         and $p$ from Equation~\ref{eqn:totflux} in columns 7, 9 and 11, along with the uncertainties $\sigma_{\mbox{\scriptsize ref}}$ and $\sigma_{\mbox{\scriptsize diff}}$
         in columns 8 and 10.
         This is an extract from the full table, which is available with the electronic version of the article (see Supporting Information).
         }
\centering
\begin{tabular}{ccccccccccc}
\hline
Variable & Filter & HJD & $M_{\mbox{\scriptsize std}}$ & $m_{\mbox{\scriptsize ins}}$ & $\sigma_{m}$ & $f_{\mbox{\scriptsize ref}}$ & $\sigma_{\mbox{\scriptsize ref}}$ & $f_{\mbox{\scriptsize diff}}$ &
$\sigma_{\mbox{\scriptsize diff}}$ & $p$ \\
Star ID  &        & (d) & (mag)                        & (mag)                        & (mag)        & (ADU s$^{-1}$)               & (ADU s$^{-1}$)                    & (ADU s$^{-1}$)                &
(ADU s$^{-1}$)                     &     \\
\hline
V1 & $V$ & 2454939.20002    & 17.167 & 18.487 & 0.003  & 661.567 & 1.722  & -299.031 & 1.098  & 1.1564 \\
V1 & $V$ & 2454939.23093    & 17.187 & 18.507 & 0.003  & 661.567 & 1.722  & -309.195 & 1.076  & 1.1625 \\
\vdots   & \vdots & \vdots  & \vdots & \vdots & \vdots & \vdots   & \vdots & \vdots   & \vdots & \vdots \\
V1 & $I$ & 2454939.19220    &16.541  & 17.935 & 0.003  & 782.781  & 3.855  &-134.485  & 2.042 & 1.1907 \\
V1 & $I$ & 2454939.20744    &16.528  & 17.922 & 0.003  & 782.781  & 3.855  &-122.903  & 1.961 & 1.1757 \\
\vdots   & \vdots & \vdots  & \vdots & \vdots & \vdots & \vdots   & \vdots & \vdots   & \vdots & \vdots \\
\hline
\end{tabular}
\label{tab:vri_phot}
\end{table*}

\subsection{Photometric Calibrations}

The instrumental $v$ and $i$ magnitudes were converted to the Johnson-Kron-Cousins photometric system by 
using the standard stars in the field of NGC 5024 from the collection of Stetson (2000)\footnote{
(http://www3.cadc-ccda.hia-iha.nrc-cnrc.gc.ca/community/STETSON/standards)}.
These stars have magnitudes in the interval $13.0 < V < 17.6$. We identified 48 and 46 standard stars in the 
$V$ and $I$ reference images with a homogeneous colour distribution between $0.04 < V-I < 1.37$. The transformations between the instrumental and the standard systems 
can be fitted by a linear relation and no colour term was found to be significant. The linear correlation coefficients are of the order of $\sim0.99$ (see Fig. \ref{transVI}).
These relations were used to transform all detected point sources in our images to the Johnson-Kron-Cousins photometric system (Landolt 1992).
All our $V,I$ photometry for the variables in the field of our images for NGC 5024 is reported in Table \ref{tab:vri_phot}.
Only a small portion of this table is given in the printed version of this paper but the full table is available in electronic form.

\subsection{Astrometry}
\label{sec:astrometry}

A linear astrometric solution was derived for the $V$ filter reference image by matching $\sim$1000 hand-picked stars with the USNO-B1.0 star catalogue
(Monet et al. 2003) using a field overlay in the image display tool {\tt GAIA} (Draper 2000). We achieved a radial RMS scatter in the residuals of
$\sim$0.3~arcsec. The astrometric fit was then used to calculate the J2000.0 celestial coordinates for all the confirmed variables in our field of view
(see Table~\ref{variables}). The coordinates correspond to the epoch of the $V$ reference image, which is the heliocentric Julian day $\sim$2455249.332~d.

\begin{figure*} 
\includegraphics[width=16.cm,height=22.cm]{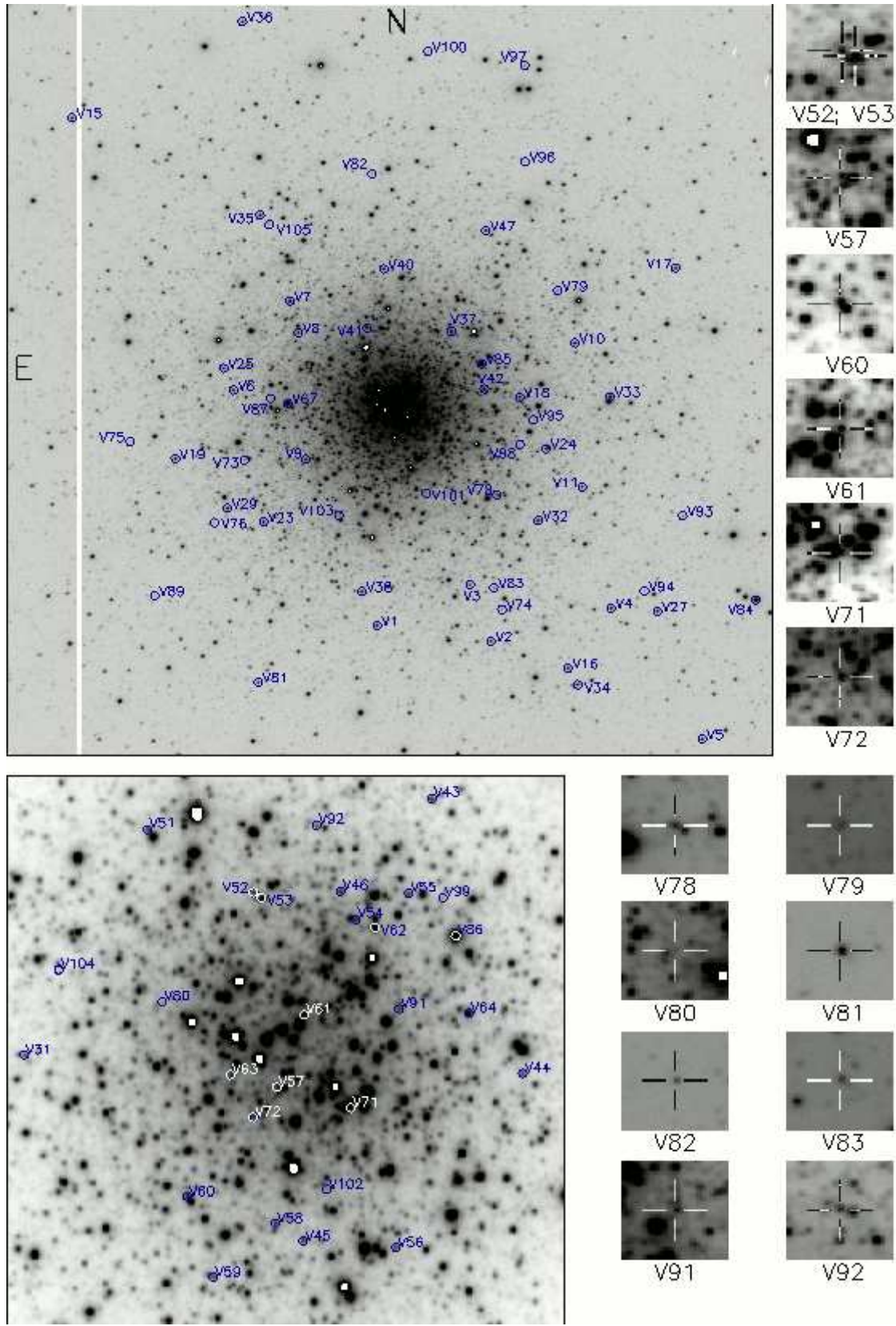}
\caption{Identification of the variable stars in the field of NGC~5024. The bottom chart corresponds 
to the central region of the cluster. Small stamps of size 14.8$\arcsec$ by 14.8$\arcsec$ are given for some RR Lyrae stars of difficult identification.  
In all images the North is up and East is to the left.}
    \label{N5024a}
\end{figure*}

\section{Variable stars in NGC 5024}

Of the 90 stars listed in the most recent version of the Catalogue of Variable stars in Globular Clusters 
(Clement et al. 2001), 13 stars are not in the field of our images and 7 are saturated. The star V39 
has been shown to be non variable by Cuffey et al. (1965). We confirm from our present data that the star is not variable at the level of 0.01 mag and hence it is not included. The remaining 69 stars are listed in 
Table \ref{variables}. In this paper we also report the discovery of two RR1 stars near the central regions of the cluster, 
now labeled V91 and V92 (included in Table \ref{variables}), and of 13 SX Phe stars. In this section we discuss the results of our exploration of their variability, periodicity, field identification and
celestial coordinates. 

\setcounter{figure}{2}
\begin{figure*} 
\includegraphics[width=18.cm,height=22.cm]{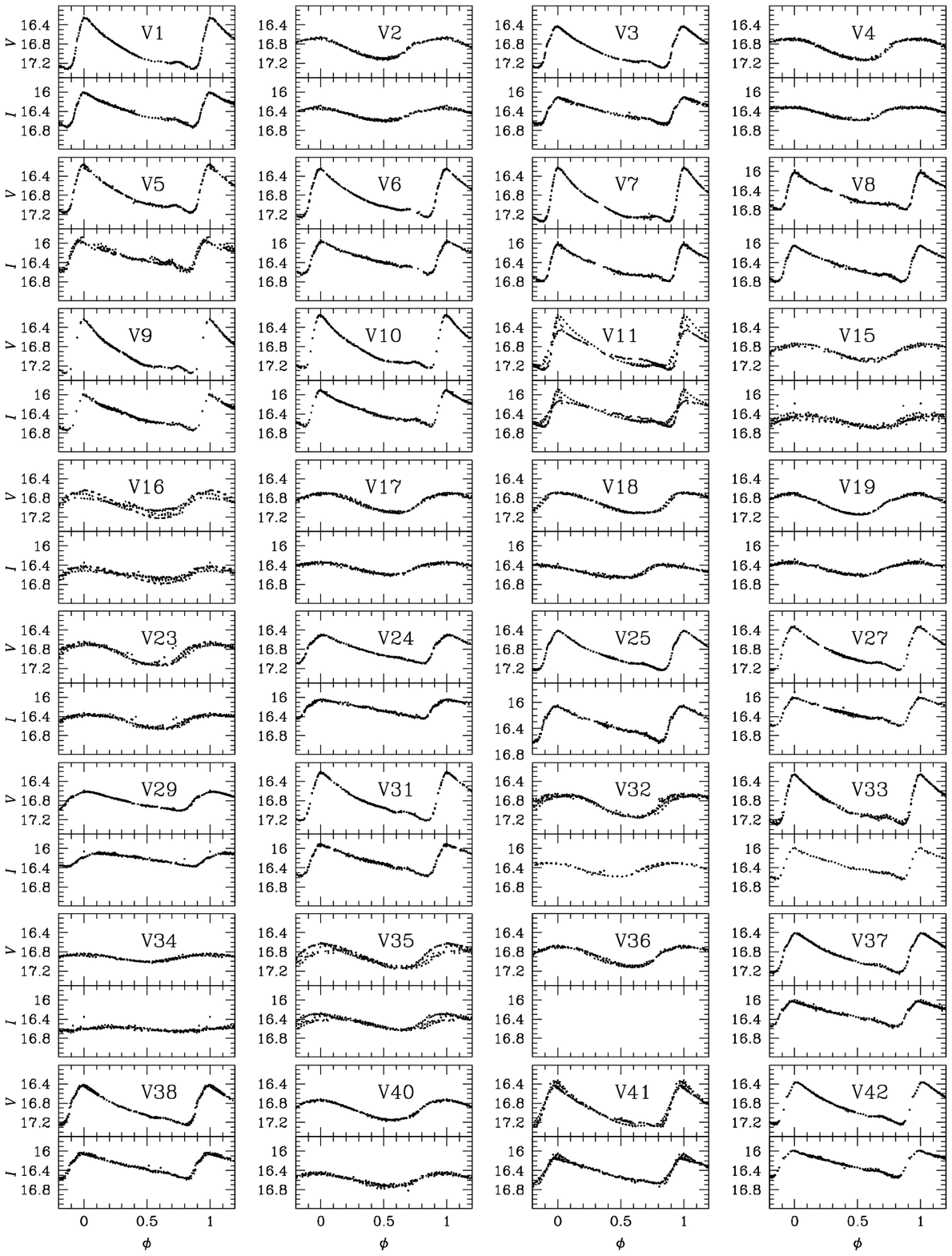}
\caption{Standard $V$ and $I$ light curves of RR Lyrae stars in NGC 5024. Star V36 falls near the edge of our images therefore it was not included in the $I$ frames, hence no $I$ light curve is available.}
    \label{VARS1}
\end{figure*}

\setcounter{figure}{2}
\begin{figure*} 
\includegraphics[width=18.cm,height=18.cm]{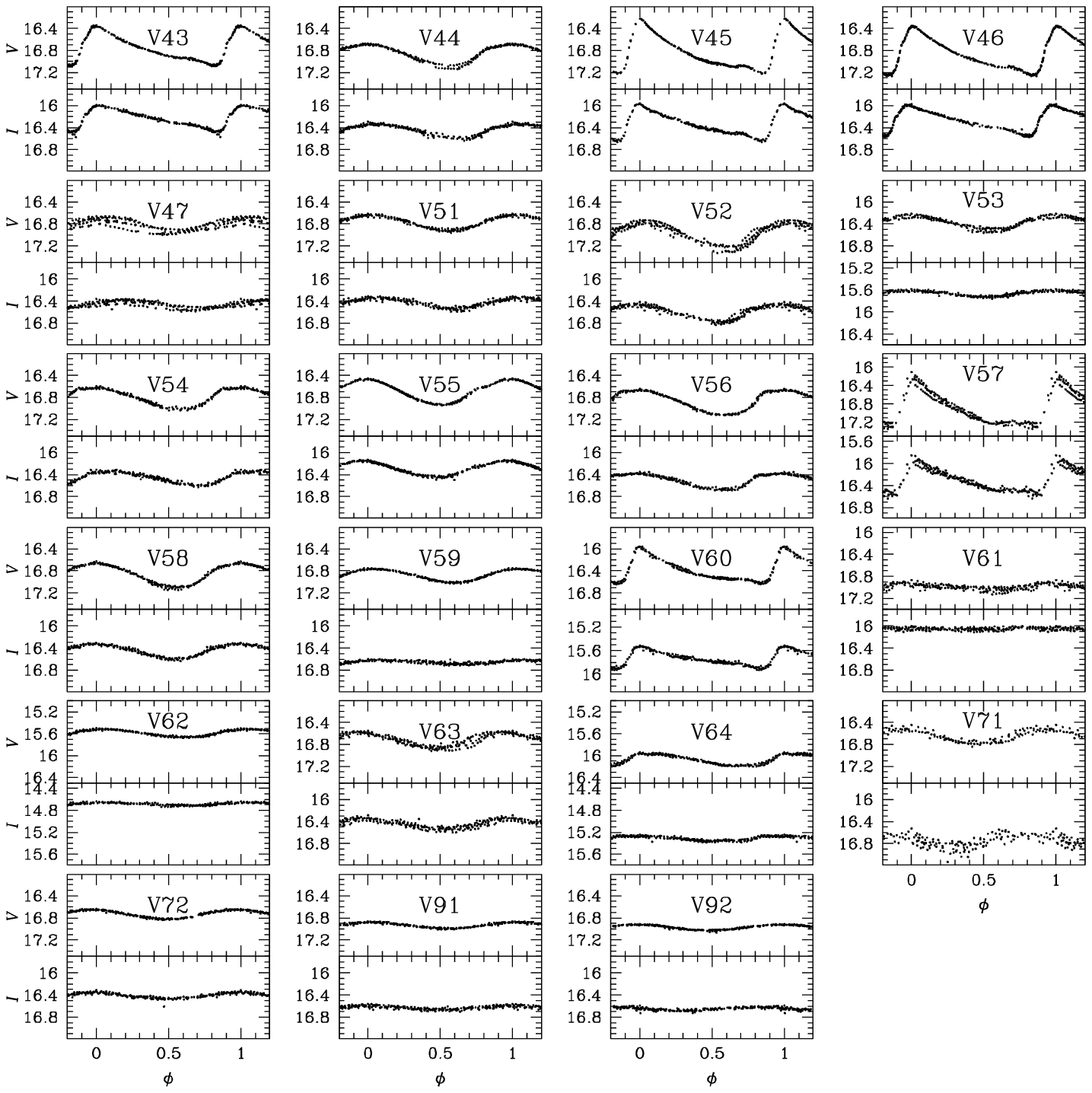}
\caption{Continued}
\end{figure*}

\subsection{Identification of variable stars}
\label{sec:IDVAR}

Many variable stars in this cluster are located in the central crowded
region and some are blended. Probably due to this the identifications 
of some variables in previous publications has been in error. Also, for many of the most
recently discovered variables no identification charts have ever been published.
The 8 candidate RR Lyrae stars announced by Rey et al. (1998) turned out to be previously 
known variables as has been pointed out by K00 and confirmed by us. 
In this paper we provide a chart and accurate
celestial coordinates for all the 
known and the newly discovered variables in the cluster. In the following paragraphs we remark 
on those variables whose identification was wrong or dubious in previous publications. 

\begin{figure} 
\includegraphics[width=8.5cm,height=12.cm]{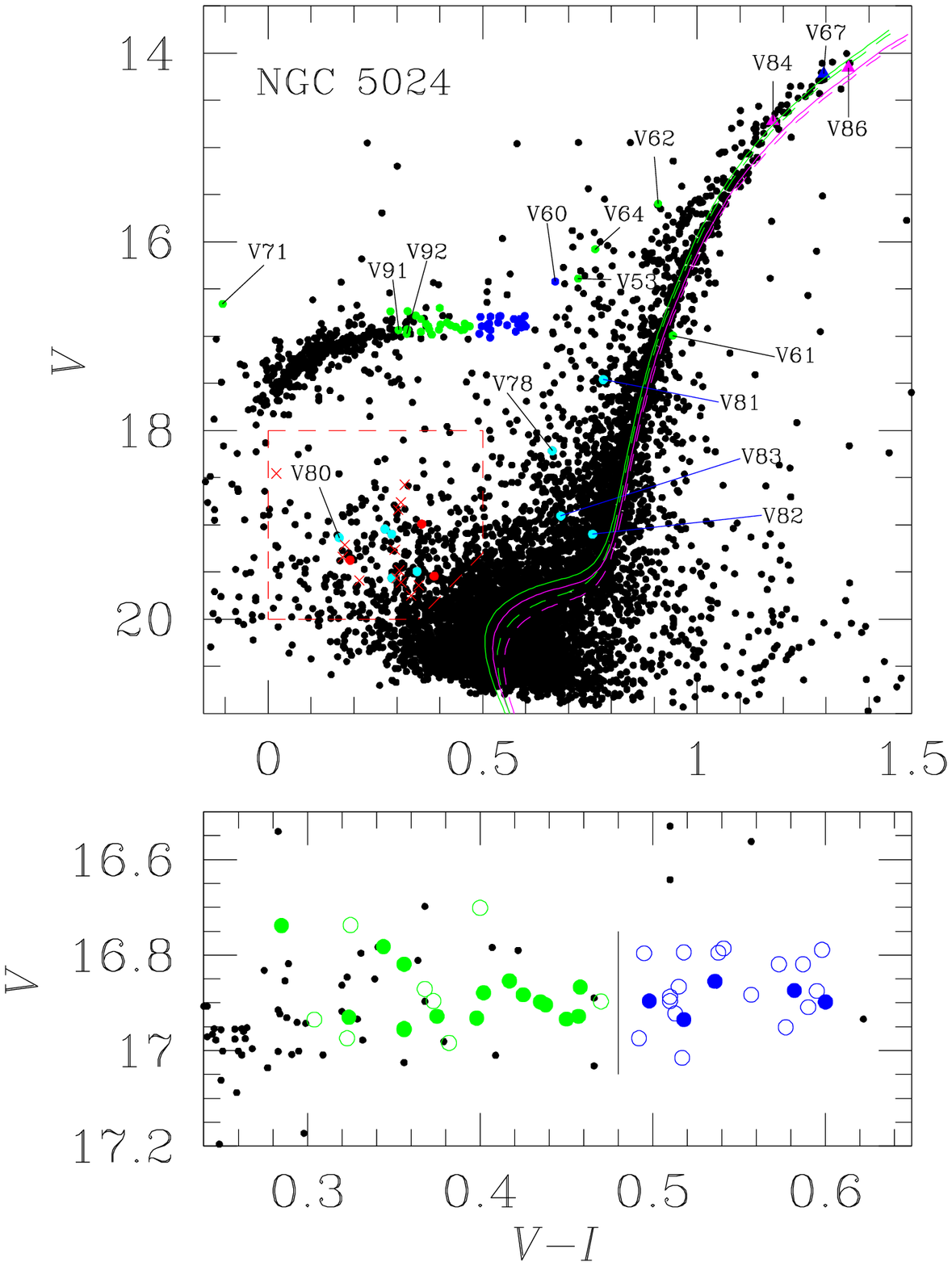}
\caption{CMD of NGC 5024. Top panel: RR0 are plotted as blue circles,
RR1 as green circles, confirmed SX Phe are plotted as red circles and suspected SX Phe as turquoise circles. The 13 newly found SX Phe stars (section \ref{sec:SXPHE}) are indicated as red crosses. The semi-regular variable V67 and the two slow irregular variables 
V84 and V86 are also labeled. A discussion on the peculiarly placed stars is given in section \ref{sec:IDVAR}. 
The isochrones are from VandenBerg, Bergbusch \& Dowler (2006) for [Fe/H]=$-1.84$ (purple) $-2.01$ (green) and 
[$\alpha$/Fe]=+0.3 for the ages 12 (continuous) and 14 (segmented) Gyrs,
see section \ref{sec:age} for a discussion. Bottom panel: a blow up of the HB region. Filled circles are stars 
with Blazhko effect, empty circles are for non-modulated stars. RR0 and RR1 stars are clearly separated and 
the border is sketched by the vertical line. Note that the RR1 with Blazhko tend to concentrate on the FO-F 
region. See section \ref{sec:CMD} for a discussion.}
\label{CMD}
\end{figure}

\noindent{\bf V57}. The star identified on the map of K00 shows no variations in
our photometry, while the star to the NE at RA=13$^{h}$ 12$^{m}$ 55.56$^{s}$ dec=+18$^{o}$ 09' 58.2'' shows a clear variation with
an amplitude of $\sim$0.8 mag and a period of $\sim$0.568198d, thus, we confirm our identification 
of V57 (see Fig. \ref{N5024a}) and its RR0 nature. Incidentally, the correct star does not appear in the chart of K00.

\noindent{\bf V60}. This star is blended with another star of similar brightness. We have confirmed that the 
variable is the star to the NE as identified on Fig. \ref{N5024a}, which coincides well
with the identification made by K00. However caution is advised on future work. Due to the
contamination of the neighbouring star, the magnitude, colour and amplitude of V60 have been
altered. Its mean $V$ and $I$ magnitudes are brighter compared to other
RR Lyraes, hence its peculiar position on the CMD of Fig. \ref{CMD}.

\noindent{\bf V61}. This star is misidentified in the chart of K00. The star marked is a $V \sim 14.3$ magnitude star and hence it is nearly three magnitudes brighter than the RR Lyrae stars 
in the cluster. The RR1 star is in fact the fainter star to the NW as marked in the 
finding chart of Fig. \ref{N5024a}. This star has a period of 0.369917 d and a light 
curve typical of an RR1 both in $V$ and $I$, thus we confirm 
the identification of the star. However due to the contamination of the brighter 
neighbour, V61, like V60, has a peculiar magnitude, color and amplitude.

\noindent{\bf V62}. This star is too bright both in $V$ and $I$ to be a RR Lyrae, although its period and light curve are those of a typical RR1. Its position on the CMD is correspondingly peculiar.
In fact in our images this is an exceptionally close blend of two bright stars as can be seen by the residuals on the reference image when the PSF model is subtracted. This explains the very low amplitude of the light curve and why it is so bright on the CMD. It is not possible to isolate the RR Lyrae star with the present data.

\noindent{\bf V64}. Similar to V62 this star in our images is a very close blend of two bright stars,
one of them being the RR1 variable. We cannot isolate the true variable. Its low amplitude and peculiar position on the
CMD are due to the blending.

\noindent{\bf V71}. We identified this star on our astrometric image using the RA and DEC from
DK09. Their coordinates on our astrometric image fall on a 14 mag star which is not variable. A close inspection of the 
neighbouring stars in our difference images led us to infer that the
variable star is the faint one blended to the W. See the detailed identification chart in
Fig. \ref{N5024a}. We confirm its variability as reported by DK09 but have found a slightly different period. 
The light curve is scattered and the colour is peculiar probably due
to contamination of two close neighbours.

\noindent{\bf V72}. We identified this star on our astrometric image using the RA and Dec from DK09.  The star marked by DK09 is not variable. We explored the neighbouring stars and found that the star labeled as V72 in Fig. \ref{N5024a} is the true RR1
which coincides in amplitude and period with the star discovered by DK09 which indicates that we have identified the genuine
variable.

\begin{figure*} 
\includegraphics[width=15.cm,height=18.cm]{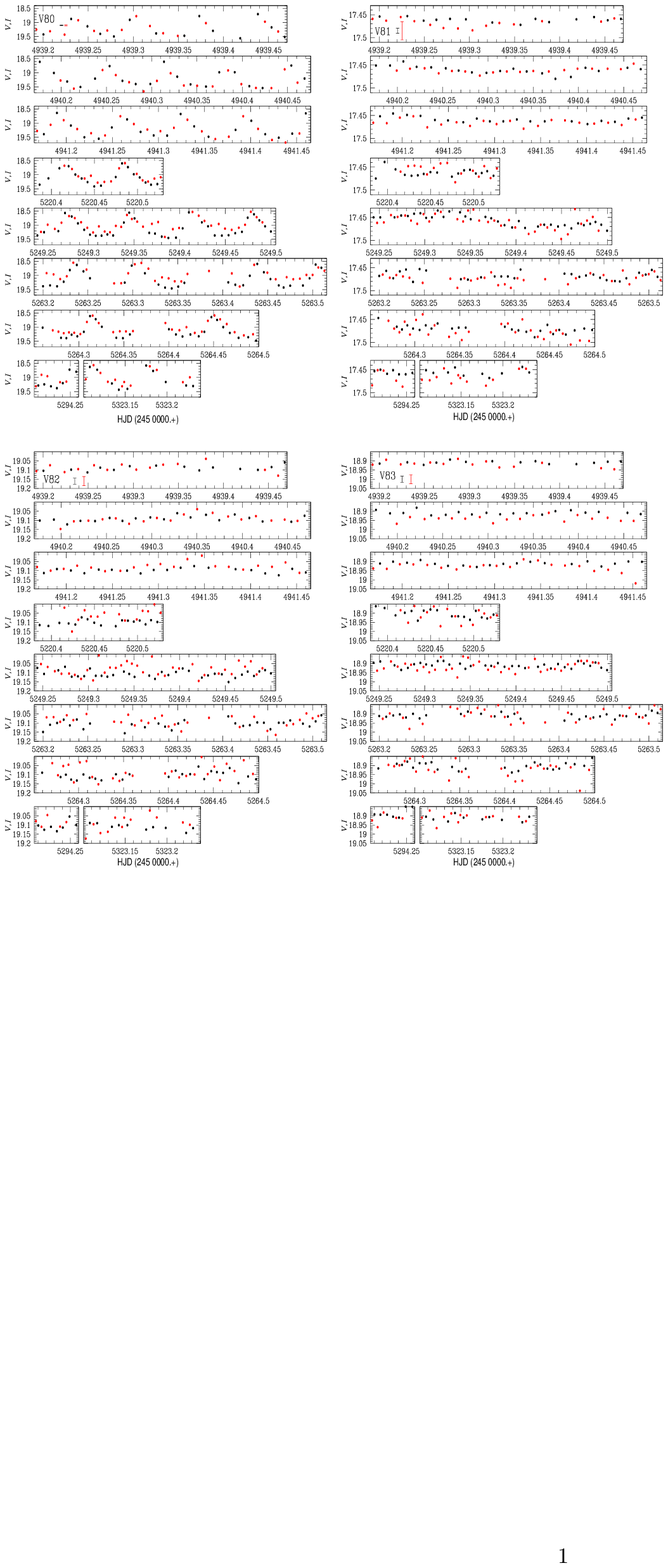}
\caption{Light curves of V80, V81, V82 and V83 suspected as SX Phe stars by DK09. Black and red symbols correspond to $V$ an $I$ data respectively. Typical mean uncertainties in $V$ and 
$I$ are indicated by the black and red error bars respectively in the top boxes. For V80 the uncertainties are of about the same size as the symbols. We only
confirm the variability and SX Phe nature of V80.}
    \label{4SXd}
\end{figure*}

\noindent{\bf V80}. This star was listed as suspected SX Phe variable by DK09  although they 
do not provide a finding chart for the star. We have used the coordinates given by these authors (RA~$\alpha = 13^{\mbox{\scriptsize h}} 12^{\mbox{\scriptsize m}} 57.4^{\mbox{\scriptsize s}}$,
Dec.~$\delta = +18\degr 10\arcmin 13.8\arcsec$) to identify the star. Their coordinates correspond to a V=17.8 magnitude star that does not show
variations above the mean photometric error for its brightness which is of about 0.01 mag. We explored the light curves of the neighbouring stars and confirm the SX Phe is the faint star to the W, at RA~$\alpha = 13^{\mbox{\scriptsize h}} 12^{\mbox{\scriptsize m}} 57.4^{\mbox{\scriptsize s}}$,
Dec.~$\delta = +18\degr 10\arcmin 15.2\arcsec$. Its light curve is shown in Fig. \ref{4SXd} and it is identified in Fig. \ref{N5024a}. 

\noindent{\bf V81}. Like DK09 we find this star to be located in the lower part of the RGB (see Fig. \ref{CMD}). We found no clear traces of variability in the light
curve of this star above the mean photometric error, which for its brightness is of about 0.01 mag (see Fig. \ref{4SXd}), and we did not succeed in finding any frequency above the noise level in the 
period spectrum. Thus we do not confirm the period found by DK09 (0.07137d) nor
we confirm this star as an SX Phe variable. 

\noindent{\bf V82 and V83}. We have identified these stars using the coordinates published by DK09. 
The stars are quite isolated, still we do not see
clear signs of variability in their light curves above the photometric error of $\sim 0.03$ mag for 
their corresponding brightness (Fig. \ref{4SXd}). The period analysis of V82 does not reproduce the frequency 
reported by DK09 (45.33$d^{-1}$). Thus we do not confirm its variability and hence its SX Phe nature. 
For V83 we find
some signal at $f_1 = 4.33$ although of very low amplitude (0.008 mag). The reported
frequency by DK09 is $f_1 = 8.01$. Thus even if some indications of variability 
might be present for V83, its status as an SX Phe variable remains uncertain.

\noindent{\bf V91 and V92}. These are two newly discovered RR1 stars with periods 0.302423d and 
0.277219d respectively. 
They are found to be located right in the HB along with the other RR1 stars known in the cluster. 
Detailed identification of these stars is given in Fig. \ref{N5024a}.

The light curves of all RR Lyrae stars in the field of our images are shown in Fig. \ref{VARS1}.

\begin{table*}
\scriptsize
\caption{General data for confirmed variables in NGC~5024.}
\label{variables}
\centering
\begin{tabular}{llllllllll}
\hline
 Variable & Bailey's & Other & $P$ (days) & HJD$_{max}$ &$P$ (days) &$P$ (days)    & $P$ (days)     & RA         & Dec. \\
          & type     & type  &this work &(+240~0000.) &J03 & K00       & DK09       & J(2000.0) &J(2000.0) \\
\hline
V1 & RR0&&0.609823&54939.438 &&-- &0.6098298 &13 12 56.32 &+18 07 13.9\\
V2 & RR1-Bl&&0.386133&54940.342 &&-- &0.386122 &13 12 50.28 &+18 07 00.9\\
V3 & RR0&&0.630598&55323.198 &&-- &0.630605 &13 12 51.38 &+18 07 45.4\\
V4 & RR1&&0.385635&54939.278 &&-- &0.385545 &13 12 43.88 &+18 07 26.4\\
V5 & RR0&&0.639421&55249.474 &&-- &0.639426 &13 12 39.08 &+18 05 42.7\\
V6 & RR0&&0.664017&55294.248 &&0.6640142 &0.664020 &13 13 03.90 &+18 10 19.9\\
V7 & RR0&&0.544861&54940.391 &&0.5448460 &0.5448584 &13 13 00.86 &+18 11 30.0\\
V8 & RR0&&0.615524&55249.468 &&0.6155072 &0.615528 &13 13 00.41 &+18 11 05.1\\
V9 & RR0&&0.600375&54941.190 &&0.6003482 &0.6003690 &13 13 00.07 &+18 09 25.1\\
V10& RR0&&0.608264&55294.226 &&0.6082530 &0.6082612 &13 12 45.72 &+18 10 55.5\\
V11& RR0-Bl&&0.629959&55220.523 &&0.6299424 &0.629940 &13 12 45.38 &+18 09 02.0\\
V15& RR1-Bl:&&0.308704&55220.482 &&-- &0.3086646 &13 13 12.38 &+18 13 55.3\\
V16& RR1-Bl&&0.303158&55294.257 &&-- &0.3031686 &13 12 46.19 &+18 06 39.3\\
V17& RR1-Bl&&0.381055&55323.183 &&-- &0.381282 &13 12 40.34 &+18 11 54.1\\
V18& RR1-Bl&&0.336050&54939.325 &&0.33611&0.336054 &13 12 48.6&+18 10 13.2\\
V19& RR1-Bl&&0.391160&55294.248 & & 0.3909871 &0.391377 &13 13 07.01 &+18 09 26.4\\
V23& RR1-Bl&&0.366099&54940.391 &&0.36579 &0.365804 &13 13 02.34 &+18 08 36.0\\
V24& RR0&&0.763195&54940.407 &&0.7631901 &0.763198 &13 12 47.28 &+18 09 32.4\\
V25& RR0&&0.705144&55264.341 &&0.7051473 &0.705162 &13 13 04.41 &+18 10 37.3\\
V27& RR0&&0.671073&54940.342 && &0.671071 &13 12 41.43 &+18 07 23.8\\
V29& RR0&&0.823243$^a$&55220.397 &&0.8232550 &0.823243 &13 13 04.26 &+18 08 47.0\\
V31& RR0&&0.705671&55323.183 &&0.70572 &0.705665 &13 12 59.57 &+18 10 04.9\\
V32& RR1-Bl&&0.390524&55323.183 &&0.39041 &0.390623 &13 12 47.73 &+18 08 35.9\\
V33& RR0-Bl&&0.624584&54941.416 &&0.624585 &0.6245815 &13 12 43.86 &+18 10 13.1\\
V34& RR1-Bl&&0.289626&54941.266 &&-- &0.289611 &13 12 45.70 &+18 06 26.2\\
V35& RR1-Bl&&0.372668&55264.280 &&-- &0.372666 &13 13 02.42 &+18 12 37.9\\
V36& RR1-Bl&&0.373320 &54941.430&&-- &0.373242 &13 13 03.29 &+18 15 10.4\\
V37& RR0&&0.717620&54941.309 & &0.717611 &0.717615 &13 12 52.28 &+18 11 05.4\\
V38& RR0-Bl&&0.705798&55220.482 &&-- &0.705792 &13 12 57.14 &+18 07 40.6\\
V40& RR1-Bl:&&0.314819&55264.433 & &0.31466 &0.3147939 &13 12 55.86 &+18 11 54.7\\
V41& RR0-Bl&&0.614429&55323.107 & &0.61455 & 0.614438&13 12 56.75 &+18 11 08.5\\
V42& RR0&&0.713713&54939.419 & &0.713694 &0.713717 &13 12 50.54 &+18 10 20.1\\
V43& RR0&&0.712008&54941.204 & &0.71205 &0.712017 &13 12 53.08 &+18 19 55.5\\
V44& RR1-Bl&&0.374924&55220.488 & &0.27276 &0.375099 &13 12 51.66 &+18 10 00.6\\
V45& RR0&&0.654946&54939.200 &&0.654966 &0.654950 &13 12 55.16 &+18 09 27.4\\
V46& RR0&&0.703649&55263.430 &&0.70363 &0.703655 &13 12 54.53 &+18 10 37.0\\
V47& RR1-Bl&&0.335377$^a$&55249.380 &&-- &0.335377&13 12 50.42 &+18 12 24.7\\
V51& RR1-Bl&&0.355216&55249.403 &&0.35519 &0.355203 &13 12 57.59 &+18 10 49.7\\
V52& RR1-Bl&&0.374357&55249.410 &&0.37414 &-- &13 12 55.92 &+18 10 37.1\\
V53& RR1-Bl&&0.389076&54941.266 &&0.38911 &-- &--&--\\
V54& RR1-Bl:&&0.315105&55323.107 &&0.31512 &0.315122 &13 12 54.31 &+18 10 31.5\\
V55& RR1&&0.443264&54939.356 &&0.30677 &0.443386 &13 12 53.46 &+18 10 36.6\\
V56& RR1&&0.328883&55220.457 &&0.32895 &0.328796 &13 12 53.69 &+18 09 26.0\\
V57& RR0-Bl&&0.568198&54941.266 &&0.56831 &0.568234 &13 12 55.56 &+18 09 58.2\\
V58& RR1-Bl&&0.354966&55263.200 &&0.35499 &0.354954 &13 12 55.60 &+18 09 31.0\\
V59& RR1&&0.303936&55263.422 &&0.30393 &0.303941 &13 12 56.66 &+18 09 20.8\\
V60& RR0&&0.644766&55294.233 &&0.64475 &0.644755 &13 12 56.98 &+18 09 36.5\\
V61& RR1&&0.369917&54940.178 &&0.37951 &-- &13 12 55.12 &+18 10 12.5\\
V62& RR1&&0.359911&55263.438 &&0.35992 &0.359891 &13 12 54.00 &+18 10 29.8\\
V63& RR1-Bl&&0.310497&55264.357 &&0.310476&0.310476 &13 12 56.29 &+18 10 00.7\\
V64& RR1-Bl&&0.319719&55249.424& &0.31955 &0.319529 &13 12 52.52 &+18 10 12.5 \\
V67&    & SR?& 29.4 &&--&-- & &13 13 01.01 &+18 10 09.6\\
V71& RR1-Bl:&&0.304508&54939.249 &&--      &0.304242&13 12 54.38 &+18 09 54.0\\
V72& RR1&&0.340749&55249.327 &&--      &0.254155&13 12 55.94 &+18 09 52.3\\
 V73& & SX&0.07011 &&0.0701 & &0.07010&13 13 03.35 &+18 09 25.1\\
 V74& & SX&0.04537 &&0.0454&  &0.04537&13 12 49.7 &+18 07 26.0\\
 V75& & SX&0.04425 &&0.0442 & &0.04425&13 13 09.4 &+18 09 39.8\\
 V76& & SX&0.04149 &&0.0415 & &0.04149&13 13 04.97 &+18 08 35.8\\
V78& & SX&0.04493 && & &0.04493&13 12 49.91 &+18 08 56.6\\
V79& & SX&0.04632 && & &0.04631&13 12 46.60 &+18 11 37.0\\
V80& & SX?&0.06743&& &-- &0.06743&13 12 57.37 &+18 10 15.2\\
V81& & SX?& --    &&--&-- &0.07137&13 13 02.69 &+18 06 29.7\\
V82& & SX?&--&& &-- &0.02206&13 12 56.46 &+18 13 09.9\\
V83& & SX?&--&&--&-- &0.12470&13 12 50.11 &+18 07 43.0\\
V84& & Lb?& --     && & &22.4  &13 12 36.17 &+18 07 32.2\\
V85& & Lb?& --     && & &19.8  &13 12 50.69 &+18 10 39.6\\
V86& & Lb?& --     && & &22.2  &13 12 52.71 &+18 10 28.1\\ 
 V87& & SX &0.04633 &&0.0479 & &--  &13 13 01.92  &+18 10 13.2\\
 V89& & SX &0.04336 &&0.0435 & &--  &13 13 08.16  &+18 07 38.5\\
V91& RR1&&0.302423& 55220.435 &&--&-- &13 12 53.62 &+18 10 13.6\\
V92& RR1&&0.277219& 54940.191 &&--&-- &13 12 54.91 &+18 10 50.3\\
\hline
\end{tabular}

\raggedright
\flushleft{\quad $^{a}$ adopted from DK09.\\ 
\quad J03: Jeon et al. (2003).}
\end{table*}

\subsection{The Colour-Magnitude-Diagram}
\label{sec:CMD}

We calculated a magnitude weighted mean to the light curves of about 8900 stars measured in the $V$ 
and $I$ frames to construct the Colour-Magnitude Diagram (CMD) shown in the top panel of
Fig. \ref{CMD}. For the non-variable stars and the sinusoidal symmetric and well phased covered
variables, like RR1 and SX Phe, the magnitude and colour weighted means are virtually identical to 
better approaches to the static star, like the intensity weighted means (Bono et al. 1995). However 
for the asymmetric RR0 stars, and despite their evenly good phase coverage, we have 
calculated $V-I$ from the mean values of $V$ and $I$ obtained from the Fourier fit of the light 
curve ($A_0$ in eq. \ref{eq_foufit}). For the five RR0 Blazhko variables we used only the light 
curve at maximum amplitude. Marked on the figure are all the 
variables in NGC 5024 contained in the field of our images and an approximate location of 
the Blue Straggler region where eclipsing binaries and SX Phe variables 
are expected to be. Most of the known RR Lyrae stars and SX Phe fall in the
expected region except for a few labeled stars. The reasons for the
peculiar positions of these stars are related either to contamination from 
a neighbouring star or an inaccurate reference flux estimation 
in a crowded region, or both. We notice the presence of non-variable stars in the RR Lyrae instability 
strip. We have inspected their light curves and confirm their non-variable nature. These stars are mostly 
located in the crowded 
central region of the cluster and then contamination by neighbours may have rendered incorrect magnitudes 
and/or colour for some of them. See sections \ref{sec:IDVAR} and \ref{sec:SXPHE} for individual comments 
for RR Lyrae and SX Phe stars respectively.

The bottom panel of Fig. \ref{CMD} displays a blow-up of the HB region. RR Lyrae stars are indicated with 
colours as in the top panel but now the non-modulated stars are plotted with empty circles, and the 
Blazhko variables with filled circles. RR0 and RR1 stars are clearly separated
by the vertical line at $V-I = 0.48$ which may be interpreted as the border between the 
inter-mode region first overtone - fundamental (OF) region and the fundamental (F) region.
As schematically shown by Caputo et al. (1978) (their Fig 3), one could only find a clean 
separation between RR1 and RR0 if the evolution on the HB is predominantly to the red.

We note that the Blazhko RR1 stars tend to clump in the OF and a detaliled discussion on the 
origin of their amplitude and phase instability being associated with their evolution will be 
presented in a forthcoming paper (Arellano Ferro et al. 2011).

\subsection{Period determination}
\label{sec:PERIOD}

To calculate the periodicity of each light curve we used the string-length method 
(Burke et al. 1970; Dworetsky 1983) which determines the best period and a corresponding normalized string-length statistic $S_{\mbox{\scriptsize Q}}$. Exploration of the light curves with the smallest values
of $S_{\mbox{\scriptsize Q}}$ by phasing them
with their best period recovered all the known RR Lyrae stars and their periods. In column
4 of Table \ref{variables} we report the period found for each RR Lyrae star and include, in columns 6 and 7, the periods given by K00 and DK09. The agreement between our periods and those of DK09 is in general very good, the differences are smaller than $\pm 50\times10^{-6}$d. Some stars having larger differences are V17, V19, V44, V63, V71 and V72. While this might be considered as an indication of a secular period change, the presence of amplitude and phase modulations (Blazhko effect), has been detected or suspected in all these stars (see Arellano Ferro et al. 2011). 
For V17, V71 and V72 we note that the periods of 
DK09 do not phase our data properly. None of these stars was
included in the photometry of K00. The data set of DK09 combined with our data have a time span of only about 3 years, thus we refrain
from a numerical analysis in search of a period change. Attention should be paid to these stars in future studies for possible period changes.

For the shorter period stars SX Phe we used the {\tt period04} 
program of Lenz \& Breger (2005) as described below.

\begin{figure} 
\includegraphics[width=8.5cm,height=7.cm]{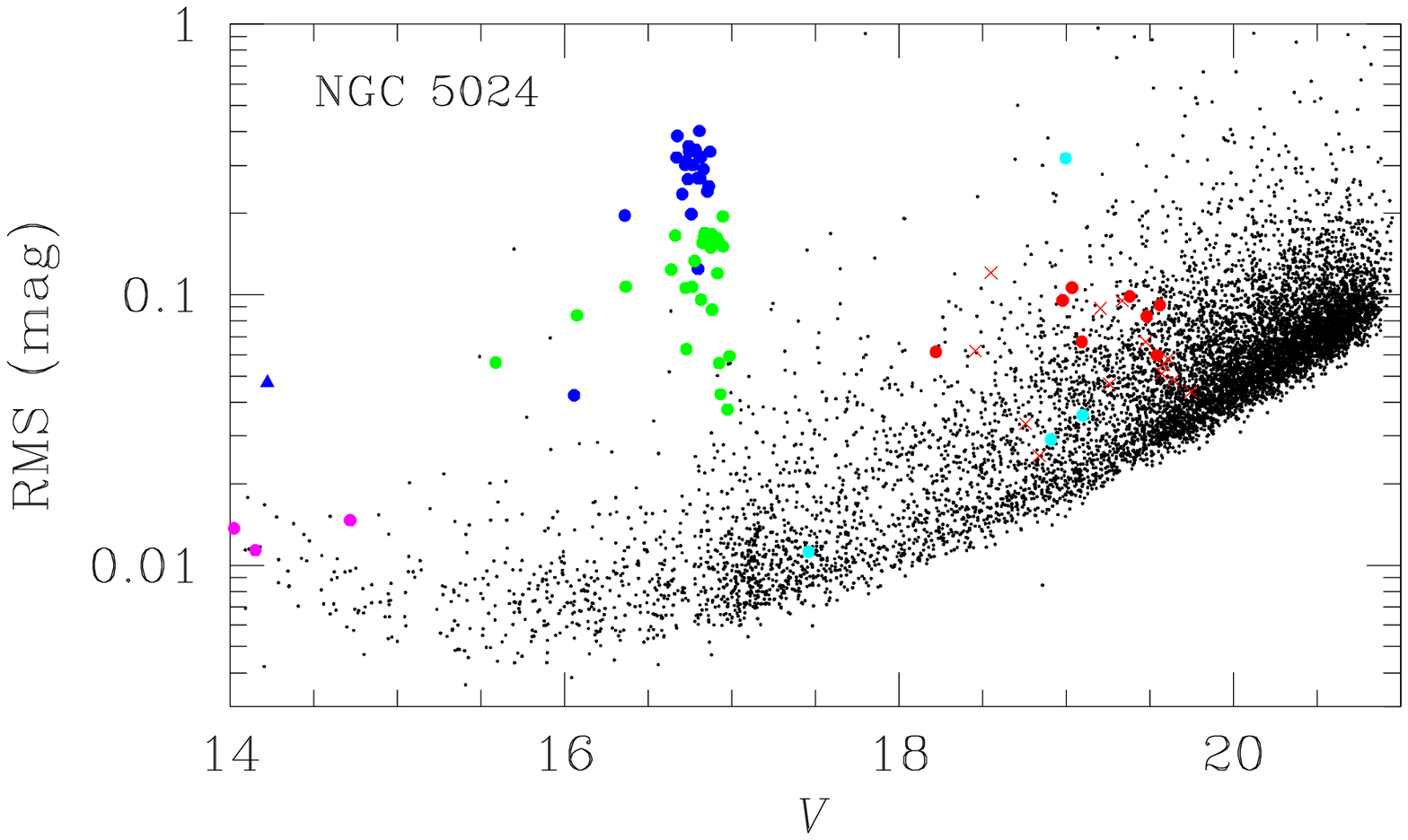}
\caption{The RMS magnitude deviations calculated for 8910 stars at 177 epochs as a function of mean magnitude $V$. 
The variable stars are indicated by colour symbols as in Fig. \ref {CMD}.}
    \label{rmsV}
\end{figure}

\subsection{Search for new variables}
\label{sec:search}

Given the deepness of our collection of images and the precision of our
photometry down to V$\sim$20, we attempted
to find new variables in the field of NGC~5024. We used four different approaches. 

First, the exploration of the statistic parameter $S_{\mbox{\scriptsize Q}}$ produced by the string-length method (section \ref{sec:PERIOD}) in stars with $S_{\mbox{\scriptsize Q}}$ smaller than a threshold value, recovered all the known RR Lyrae
and brighter variables and allowed us to discover variables V91 and V92. We note that this method fails for identifying the small amplitude and short 
period SX Phe stars, for which a different approach was used (see also Arellano Ferro et al. 2010).

Second, from Fig. \ref{rmsV} it can be seen that variable stars have larger 
values of the RMS for its corresponding mean magnitude. However, some stars
in that plot may have large RMS due to outlier photometric measurements.
We have individually explored the light curves of a large fraction of these
"candidates" down to V$\sim 17.3$ mag and with RMS larger than 0.01 mag to discriminate between authentic variables and 
false detections. We used this approach mainly to search for RR Lyrae stars
or brighter variables. For the fainter variables in the Blue Straggler region we
used a different approach. By adopting this method we confirmed the discovery of the two RR1 stars V91 and V92.

Third, mainly searching for eclipsing binaries and SX Phe stars in the 
$18 < V < 20$ mag range, the sequence of difference images in our data
collection were inspected.
By converting each difference image $D_{kij}$ to an image of absolute deviations in units of
sigma $D^{\prime}_{kij}=|D_{kij}|/\sigma_{kij}$  and then constructing the sum of all such images 
$S_{ij}=\sum_{k}D^{\prime}_{kij}$ for each filter,
one can identify candidate variable sources as PSF-like peaks in the image $S_{ij}$. Using 
this method we recovered all known the RR Lyrae and SX Phe stars, confirmed the variability of the  previously suspected SX Phe star 
V80 (DK09), and discovered 13 new SX Phe stars.

Fourth, the individual $V$ and $I$ light curves of all the stars contained in the red box in the CMD 
of Fig. \ref{CMD} were inspected. To allow for uncertainties in the $V-I$ colour and hence in 
the position of the candidate star on the CMD, stars with $(V-I) < 0.1$ were also 
checked. A total of 400 candidates were identified.
Since the $V$ and $I$ observations were carried out alternatively, in order to have a more 
continuous light curve we applied an arbitrary shift in magnitude to the $I$ data to match the $V$ data. 
Then, when variations were seen in the light curve, we performed a period analysis using  
{\tt period04}. To a certain extent the majority of stars in the BS
region present some kind of small amplitude irregular variations. However we classify a star 
as variable when 
prominent clear variations are seen above the mean photometric uncertainties and the data produced a frequency spectrum with a
peak whose signal is at least twice that of the frequency noise level. This procedure
recovered all the known SX Phe stars listed in Table \ref{variables} and confirmed the discovery of 
13 new SX Phe stars which have been labeled as V93, 
V94,..., V105. These new variables are also identified in the charts of Fig. \ref{N5024a} and 
discussed in section \ref{sec:SXPHE}.

\section{RR Lyrae stars}

\subsection{Oosterhoff type and Bailey diagram}

The average periods of the 24 RR0 and 31 RR1 stars in Table \ref{variables} are  0.661$\pm$0.064
and 0.347$\pm$0.039 d respectively and 55\% of the RR Lyrae stars in this cluster are RR1. The corresponding numbers for OoII clusters from the 
CVSGC are 0.659 and 0.368 respectively, and 48\% are RR1 stars (Clement et al. 2001). These numbers clearly identify NGC~5024 as an OoII cluster.
 
\begin{figure} 
\includegraphics[width=8.cm,height=13.5cm]{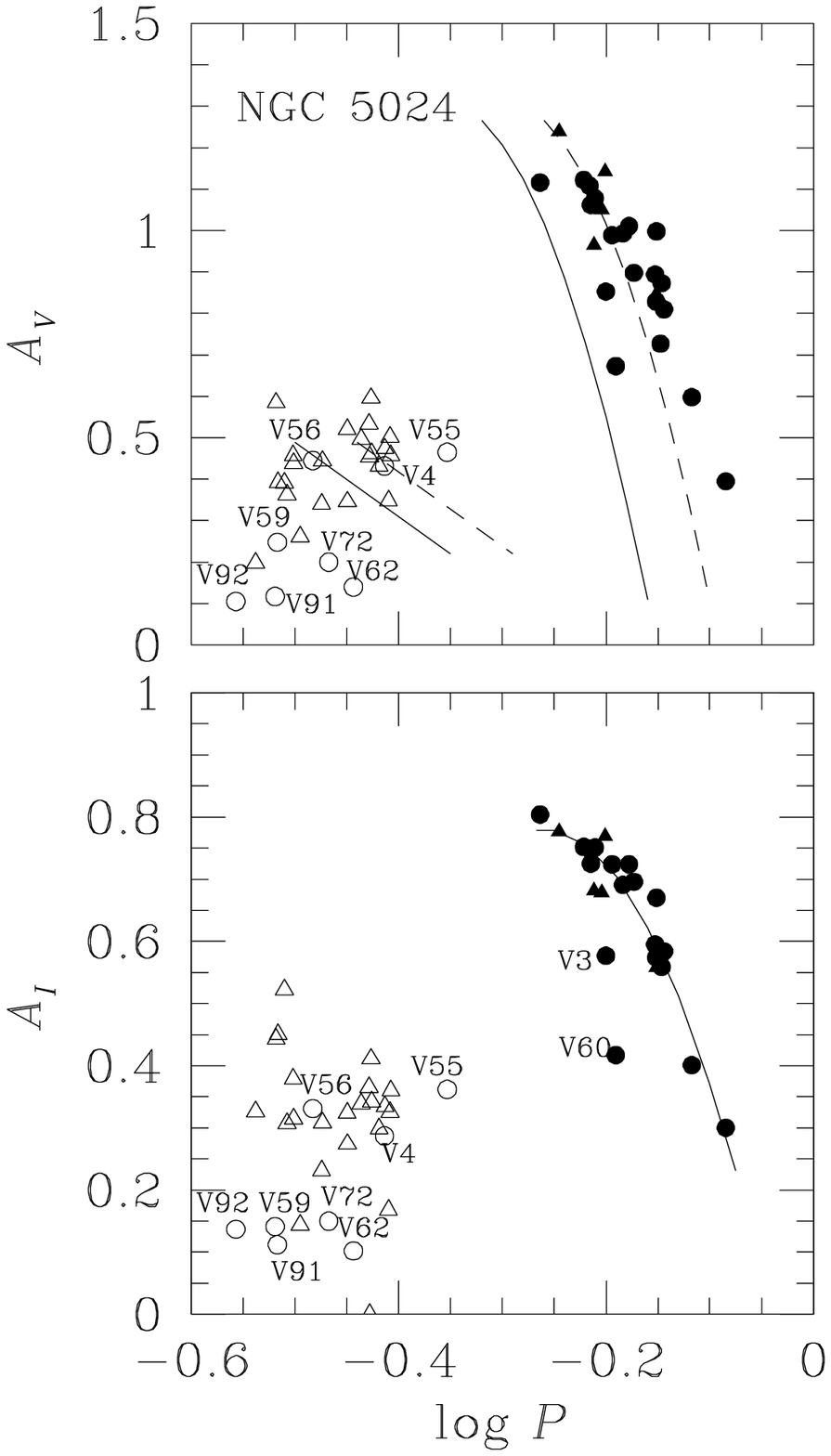}
\caption{Bailey diagram of the RR Lyrae stars in NGC 5024 in the $V$ and $I$ bands. Filled symbols represent RR0 stars and 
open symbols the RR1 stars. Circles are amplitude non-modulated stars and triangles are amplitude 
modulated or Blazhko variables. The continuous lines in the upper panel represent the average distribution of the 
RR0 and RR1 stars in M3, and the segmented lines are the loci of evolved stars (Cacciari et al. 2005). The continuous line in the bottom panel is a least squares fit to the RR0 stars (V3 and V60 have been omitted) and corresponds to eq. \ref{eq:lP_AI_OoII}.}
    \label{bailey}
\end{figure}

A plot of amplitude versus period in Fig. \ref{bailey} (the Bailey diagram) offers and additional
insight to the Oosterhoff classification and 
is useful in identifying stars with peculiar amplitudes which may introduce
noise in the determination for [Fe/H] and M$_V$ by the Fourier decomposition of their light 
curves. The top panel of Fig. \ref{bailey} shows the distribution of RR0 and RR1 stars, compared with the mean
distributions of the OoI cluster M3, shown as solid curves and taken from Cacciari et al. (2005). 
It is clear that the RR0 stars are displaced towards 
longer periods. Similar distributions are displayed by the OoII clusters
M9, M15 and M68 (see Fig. 4 of Cacciari et al. 2005). No outliers
are seen among the RR0 stars, and the five stars known to display the Blazhko effect (V11, V33,
V38, V41, V57, plotted as filled triangles) do follow the trend of the RR0 stars equally well.
This confirms the Oosterhoff type II of the cluster.

The distribution of RR1 stars is very scattered and can be seen that many appear to have shorter than expected periods for a given amplitude. A similar result was found by Contreras et al. (2010) for the OoI cluster M62 and these authors cite Clement \& Shelton (1999) as ascribing these differences to metallicity dependence, in the sense that for RR1 stars present systematic deviations toward shorter period with increasing metallicity. Since NGC~5024 is more
metal poor, this explanation is not satisfactory. We note that Clement \& Shelton based their comment on three OoI clusters with a low number of RR1 stars (M107, M4 and M5) which in fact
do not show the large scatter seen in Fig.\ref{bailey} or for M62 (Contreras et al. 2010).

It is true that Fig.\ref{bailey} contains a large number of RR1 stars with Blazhko amplitude and phase modulations (open triangles; to be discussed by Arellano Ferro et al. 2011). For these stars we have
measured the maximum observed amplitude which in fact may be underestimated if we have failed to observe the star at maximum amplitude. 
Given the large incidence of Blazhko variables among the RR1 stars we think that it is possible that even the stars with 
apparently stable light curves (open circles in 
Fig. \ref{bailey}), may turn out to show amplitude modulations if precise photometry is obtained over 
long time spans. If the Blazhko effect is a temporary peculiarity in the evolution of an RR  Lyrae star 
that may involve a switch of pulsation modes, as it seems to be the case of star V79 
in M3 (Goranskij, Christine \& Thompson, 2010), the RR1 stars in NGC~5024 may be evolving through such a
stage and further monitoring may reveal unprecedented observations of structural changes in stars in a 
globular cluster.

While the presence of Blazhko variables may contribute to the dispersion, we note that
such a dispersion does not come as a surprise on theoretical grounds. Theoretical calculations by Bono et al. (1997) show that the distribution of 
first overtone pulsators on the log P vs. $A_{bol}$ plane is not linear for a given luminosity, mass and helium content. Thus, as these quantities may have intrinsic scatter from star to star in a cluster, the resulting distribution may look as scattered as it is observed in NGC~5024 (and in M62).

As a new feature, in the bottom panel of Fig. \ref{bailey} we have produced the Bailey diagram using the RR Lyrae amplitudes 
in the $I$ filter, $A_I$. The star distribution resemble that in the $\log P- A_V$ plane. The solid line is a least squares fit to the RR0 stars (V3 and V60 were omitted) and has the form of 
eq. \ref{eq:lP_AI_OoII} which, being NGC~5024 a OoII cluster rich in RR Lyrae stars, it should 
be representative of the RR0 stars distribution in the $I$ version of the Bailey's diagram for an OoII cluster.

$$A_I = (-0.313\pm0.112) - (8.467\pm1.193) \log P$$
\begin{equation}
~~~~~~~ ~-~ (16.404\pm0.441) \log P^2.
\label{eq:lP_AI_OoII}
\end{equation}

The Oosterhoff type can also be defined from the average physical parameters of RR0 and RR1 stars derived by the Fourier decomposition of their light curves (Corwin et al. 2003). 
In the following section we shall show that the derived parameters for the RR0 stars support the OoII type for NGC~5024.

The Oosterhoff type determination of globular clusters is of relevance in early formation history of the 
Galactic halo. The scenario has been summarized most recently by Catelan (2009) who describes in detail how 
the Oosterhoff dichotomy present among the
bona fide Galactic globular clusters is not shown by clusters associated with several neighbouring galaxies, 
from which it is inferred that the halo was not formed by accretion of
dwarf galaxies similar to the present Milky Way satellites.

\begin{table*}
\caption{Fourier coefficients $A_{k}$ for $k=0,1,2,3,4$, and phases $\phi_{21}$, $\phi_{31}$ and $\phi_{41}$,
         for the 19 RR0 and 9 RR1 type variables for which the Fourier decomposition fit was successful. The numbers in parentheses indicate
         the uncertainty on the last decimal place. Also listed are the number of harmonics $N$ used to fit the light curve
         of each variable, and the deviation parameter $D_{\mbox{\scriptsize m}}$ (see Section~\ref{sec:FEMV}).
        }
\centering                   
\begin{tabular}{lllllllllrr}
\hline
Variable      & $A_{0}$    & $A_{1}$   & $A_{2}$   & $A_{3}$   & $A_{4}$   & $\phi_{21}$ & $\phi_{31}$ & $\phi_{41}$ & $N$   & $D_{\mbox{\scriptsize m}}$ \\
Star ID       & ($V$ mag)  & ($V$ mag) & ($V$ mag) & ($V$ mag) & ($V$ mag) &             &             &             &       &                            \\
\hline
              &            &           &           &           & RR0 stars &             &             &             &       &                            \\
\hline
V1 & 16.930(1) &0.365(2)   &0.179(2)   &0.127(2)   &0.085(2)  &3.904(11)     &8.187(16)  &6.166(23)  &10  &0.8\\
V3 & 16.956(1) &0.299(2)   &0.147(2)   &0.100(2)   &0.064(2)  &3.987(15)     &8.332(22)  &6.440(32)  &10  &1.2\\
V5 & 16.791(1) &0.338(2)   &0.165(2)   &0.114(2)   &0.081(2)  &3.903(14)     &8.196(22)  &6.204(30)  &10  &1.7\\
V6 & 16.862(1) &0.338(1)   &0.174(1)   &0.116(1)   &0.078(1)  &4.014(11)     &8.323(17)  &6.431(24)  &9   &1.4\\
V7 & 16.980(1) &0.393(2)   &0.177(2)   &0.137(2)   &0.091(2)  &3.823(13)     &7.946(18)  &5.840(28)  &10  &0.7\\
V8 & 16.973(1) &0.367(1)   &0.178(1)   &0.127(1)   &0.087(1)  &3.955(11)     &8.218(15)  &6.260(21)  &10  &0.9\\
V9 & 16.963(1) &0.384(1)   &0.182(1)   &0.138(1)   &0.089(1)  &3.865(11)     &8.099(16)  &5.962(22)  &10  &1.8\\
V10& 16.862(1) &0.383(1)   &0.180(1)   &0.134(1)   &0.090(1)  &3.886(8)      &8.100(11)  &6.023(16)  &10  &1.4\\
V24& 16.826(1) &0.223(1)   &0.104(1)   &0.057(1)   &0.025(1)  &4.328(15)     &8.891(25)  &7.265(50)  &10  &1.7\\
V25& 16.900(1) &0.286(2)   &0.145(2)   &0.094(2)   &0.048(2)  &4.165(16)     &8.481(6)   &6.748(42)  &8   &1.5\\
V27& 16.867(1) &0.310(2)   &0.158(2)   &0.104(2)   &0.063(2)  &4.079(13)     &8.421(20)  &6.622(30)  &9   &1.1\\
V29& 16.821(1) &0.167(1)   &0.061(1)   &0.029(1)   &0.010(1)  &4.384(23)     &9.077(41)  &7.908(108) &5   &4.2\\
V31& 16.809(1) &0.341(1)   &0.182(1)   &0.114(1)   &0.069(1)  &4.147(1)      &8.495(15)  &6.802(22)  &7   &2.8\\
V37& 16.880(1) &0.289(2)   &0.147(2)   &0.090(2)   &0.045(2)  &4.212(15)     &8.652(23)  &7.028(40)  &10  &2.1\\
V42& 16.870(1) &0.305(1)   &0.153(1)   &0.097(1)   &0.055(1)  &4.161(12)     &8.558(18)  &6.857(29)  &7   &1.9\\
V43& 16.771(1) &0.257(2)   &0.129(2)   &0.080(2)   &0.040(2)  &4.184(16)     &8.588(25)  &6.863(43)  &7   &1.1\\
V45& 16.848(1) &0.337(1)   &0.171(1)   &0.116(1)   &0.080(1)  &3.967(10)     &8.253(15)  &6.260(21)  &10  &1.1\\
V46& 16.886(1) &0.312(2)   &0.157(2)   &0.099(2)   &0.061(2)  &4.132(15)     &8.489(24)  &6.764(36)  &9   &3.0\\
V60& 16.392(1) &0.224(2)   &0.114(2)   &0.075(2)   &0.051(2)  &3.892(25)     &8.174(36)  &6.156(51)  &8   &2.3\\
\hline

             &            &           &           &           & RR1 stars &             &             &             &       & \\
\hline
V4  &16.906(1)  &0.223(2)   &0.031(2)   &0.021(2)   &0.009(2)  &4.827(58)     &3.592(85)  &1.852(187)  &7& ---   \\
V40 &16.949(1)  &0.206(1)   &0.031(1)   &0.008(1)   &0.007(1)  &4.590(38)     &2.781(136) &1.398(153)  &6& ---   \\
V55 &16.703(1)  &0.228(1)   &0.005(1)   &0.012(1)   &0.009(1)  &5.215(212)    &5.026(83)  &3.396(115)  &6& ---   \\
V56 &16.894(1)  &0.231(1)   &0.040(1)   &0.020(1)   &0.015(1)  &4.715(34)     &2.954(62)  &1.620(89)   &7& ---   \\
V59 &16.886(1)  &0.125(1)   &0.012(1)   &0.002(1)   &0.001(1)  &5.013(75)     &2.766(414) &1.838(1832) &4& ---   \\
V62 &15.592(1)  &0.074(1)   &0.009(1)   &0.004(1)   &0.002(1)  &4.506(122)    &3.117(283) &2.082(432)  &4& ---   \\
V64 &16.082(1)  &0.111(2)   &0.024(2)   &0.014(2)   &0.010(2)  &4.500(73)     &2.513(122) &1.154(175)  &4& ---   \\
V91 &16.938(1)  &0.056(1)   &0.004(1)   &0.002(1)   &--        &4.430(302)    &6.217(571) &--          &3& ---   \\
V92 &16.976(1)  &0.053(1)   &0.001(1)   &--         &--        &1.982(972)    &--         &--          &2& ---   \\
\hline
\end{tabular}
\label{tab:fourier_coeffs}
\end{table*}

NGC~5024 is indeed a rather emblematic OoII cluster in the sense that it has the second largest 
number of RR Lyrae stars, after M15, and a very blue HB ($\cal L$= (B-R)/(B+V+R) = 0.81; 
Rey et al. (1998)). This and its low metallicity places the cluster among other OoII Galactic globulars, 
following the clear Oosterhoff dichotomy found in Galactic globular clusters.

\subsection{Fourier light curves decomposition of the RR Lyrae stars}

The Fourier decomposition of RR Lyrae light curves into their harmonics has proved to be 
useful to estimate physical parameters, such as metallicity, luminosity and  temperatures through 
the use of semi-empirical relationships (e.g. Jurcksic \& Kov\'acs
1996, Morgan et al. 2007). The light curves can be represented by an equation of the form:

\begin{equation}
m(t) = A_o ~+~ \sum_{k=1}^{N}{A_k ~\cos~( {2\pi \over P}~k~(t-E) ~+~ \phi_k ) },
\label{eq_foufit}
\end{equation}

\noindent
where $m(t)$ are magnitudes at time $t$, $P$ the period and $E$ the epoch. A linear
minimization routine is used to fit the data with the Fourier series model, deriving
the best fit values of $E$ and of the amplitudes $A_k$ and phases $\phi_k$ of the sinusoidal components. 

From the amplitudes and phases of the harmonics in eq. \ref{eq_foufit}, the Fourier parameters, 
defined as $\phi_{ij} = j\phi_{i} - i\phi_{j}$, and $R_{ij} = A_{i}/A_{j}$, 
were calculated. 
The mean magnitudes $A_0$, and the Fourier light curve fitting parameters of the individual RR0 and
 RR1 type stars in
$V$ are listed in Table ~\ref{tab:fourier_coeffs}.

\subsection{[Fe/H] and M$_V$ from light curve Fourier decomposition}
\label{sec:FEMV}

The Fourier decomposition parameters can be used to calculate [Fe/H] and M$_V$ for both RR0 and RR1 stars
by means of semi empirical calibrations given in eqs. \ref{eq:JK96}, \ref{eq:ḰW01}, \ref{eq:Morgan07} and \ref{eq:K98}. 

The calibrations for [Fe/H] and M$_V$ used for RR0 stars are:

\begin{equation}
	{\rm [Fe/H]}_{J} = -5.038 ~-~ 5.394~P ~+~ 1.345~\phi^{(s)}_{31},
\label{eq:JK96}
\end{equation}

and

\begin{equation}
M_V = ~-1.876~\log~P ~-1.158~A_1 ~+0.821~A_3 + K,
\label{eq:ḰW01}
\end{equation}

\noindent given by Jurcsik \& Kov\'acs (1996) and Kov\'acs \& Walker (2001) respectively. In eq. \ref{eq:ḰW01} we have used K=0.41 to scale the
luminosities of the RR0 with the distance modulus of 18.5 for the LMC (see the discussion in Arellano Ferro et al. 2010, in
their section 4.2). The metallicity scale of eq. \ref{eq:JK96} was transformed into the widely used scale of Zinn \& West (1984)
using the relation [Fe/H]$_{J}$ = 1.431[Fe/H]$_{ZW}$ + 0.88 (Jurcsik 1995). These two metallicity scales closely coincide for [Fe/H]$\sim -2.0$ while for [Fe/H]$\sim -1.5$,
the [Fe/H]$_{J}$ is about 0.24 dex less metal poor than [Fe/H]$_{ZW}$ (see also Fig. 2 of 
Jurcsik 1995). Therefore, for a metal poor cluster such as NGC 5024, the two scales are not significantly different.

\begin{table*}
\footnotesize
\begin{center}
\caption[Parametros estelares de las RR Lyrae del tipo ab] {\small Physical parameters for the RR0 stars. The numbers in parentheses indicate the uncertainty of [Fe/H] on the last 
decimal places calculated with eq. \ref{eq:JK96} from Jurcsik \& Kov\'acs (1996).}
\label{fisicosAB}
\hspace{0.01cm}
 \begin{tabular}{lcccccc}
\hline 
Star&[Fe/H]$_{ZW}$ & $M_V$ & log$(L/L_{\odot})$ & $T_{\rm eff}$  &$D$ & $M/M_{\odot}$\\
   & & (mag)& & (K) & (kpc) &\\
\hline
V1 &$-1.692(30)$&0.50&1.702&6381.&18.81&0.71\\
V3 &$-1.634(37)$&0.52&1.691&6333.&18.81&0.68\\
V5 &$-1.796(38)$&0.48&1.709&6315.&17.80&0.70\\
V6 &$-1.769(34)$&0.45&1.721&6289.&18.64&0.71\\
V7 &$-1.674(33)$&0.56&1.675&6467.&18.67&0.72\\
V8 &$-1.685(30)$&0.49&1.706&6372.&19.29&0.71\\
V9 &$-1.739(30)$&0.49&1.702&6399.&19.11&0.71\\
V10&$-1.768(28)$&0.48&1.707&6378.&18.35&0.72\\
V24&$-1.609(58)$&0.49&1.733&6159.&18.58&0.66\\ 
V25&$-1.775(43)$&0.44&1.724&6211.&19.03&0.69\\ 
V27&$-1.702(39)$&0.46&1.715&6276.&18.57&0.69\\
V29&$-1.660(80)$&0.40&1.740&6026.&18.71&0.68\\
V31&$-1.764(37)$&0.39&1.743&6233.&18.66&0.72\\
V37&$-1.662(47)$&0.42&1.732&6216.&19.04&0.69\\ 
V42&$-1.735(41)$&0.41&1.736&6218.&19.03&0.70\\
V43&$-1.701(49)$&0.46&1.718&6209.&17.82&0.67\\
V45&$-1.800(29)$&0.46&1.716&6299.&18.42&0.70\\
V46&$-1.762(44)$&0.42&1.733&6226.&19.12&0.71\\ 
V60&$-1.836(56)$&0.57&1.672&6253.&14.19$^a$&0.65\\ 
\hline
Weighted &&&&&&\\
Mean & -1.724(9)&  0.46 & 1.714 & 6282. & 18.68& 0.70\\
$\sigma$ &$\pm$0.06 &$\pm$0.05 & $\pm$0.020 & $\pm$101. & $\pm$0.42 &$\pm$0.02\\
\hline
\end{tabular}
\end{center}
$^a$value not included in the average.
\end{table*}

For the RR1 stars we employ the calibrations: 

$$ {\rm [Fe/H]}_{ZW} = 52.466~P^2 ~-~ 30.075~P ~+~ 0.131~\phi^{(c)~2}_{31}  $$
\begin{equation}
~~~~~~~	~-~ 0.982 ~ \phi^{(c)}_{31} ~-~ 4.198~\phi^{(c)}_{31}~P ~+~ 2.424,
\label{eq:Morgan07}
\end{equation}

and

\begin{equation}
M_V = 1.061 ~-~ 0.961~P ~-~ 0.044~\phi^{(s)}_{21} ~-~ 4.447~A_4, 
\label{eq:K98}	
\end{equation}

\noindent given by  Morgan et al. (2007) and Kov\'acs (1998) respectively. For eq. \ref{eq:K98} the zero point was reduced to 1.061 to make the
luminosities of the RR1 consistent with the distance modulus of 18.5 for the LMC (see discussions by Cacciari et al. 2005 and Arellano Ferro et al. 2010). The original zero point given by Kov\'acs (1998) is 1.261.

In the above calibrations the phases are calculated either from series of sines or of cosines as indicated by the superscript. We transformed our 
cosine series phases into the sine ones where necessary by the correlation ~~ $\phi^{(s)}_{jk} = \phi^{(c)}_{jk} - (j - k) {\pi \over 2}$.

Although it has been argued that the 
[Fe/H]($P,\phi_{31}$) relationship gives good results for mean light curves in Blazhko RR0 variables (Jurcsik et al. 2002; Cacciari et al. 2005; 
Jurcsik et al. 2009), 
we choose to use only the high quality light curves of 
stars with stable light curves to reduce the scatter and uncertainties in the estimation of the metallicity. Since for the RR1 stars
no similar tests have been done, and having detected common amplitude modulations among RR1 stars in this cluster
(Arellano Ferro et al. 2011), we prefer to avoid those multiperiodic variables. 

There are 24 RR0 and 31 RR1 stars reported in Table \ref{variables}.
The Blazhko effect was found by DK09 in 2 RR0 (V11 and V57) and 1 RR1 (V16). 
We discovered the effect in a further 3 RR0 and 21 RR1 and the discussion of these stars
will be presented in a separate paper (Arellano Ferro et al. 2011). For the remaining
19 RR0 and 9 RR1, their light curves seem to be stable enough to perform a Fourier decomposition
and to calculate their physical parameters. 

From the apparently non-modulated RR1 stars
we omitted six stars with too small amplitudes for their period shown in Figs. \ref{VARS1} and 
\ref{bailey}, V59, V61, V62, V72, V91 and V92.

The results are reported in {Tables  ~\ref{fisicosAB} and {\ref{fisicosC} for all the mono periodic stars without apparent amplitude modulation.
Eq. \ref{eq:JK96} is applicable to RR0 stars with a  {\it deviation parameter} $D_m$, defined by
Jurcsik \& Kov\'acs (1996) and Kov\'acs \& Kanbur (1998), not exceeding a upper limit.
These authors suggest $D_m \leq 3.0$. The $D_m$ is listed in the last column of Table~\ref{tab:fourier_coeffs}.

The values of $M_V$ in Tables~\ref{fisicosAB} and \ref{fisicosC} were transformed into $log~L/L_\odot$. The bolometric correction was calculated using the formula $BC= 0.06$ [Fe/H]$_{ZW} + 0.06$
given by Sandage \& Cacciari (1990). We adopted the value $M_{bol}^{\odot} = 4.75.$ 

\begin{table}
\footnotesize{
\begin{center}
\caption[Parametros estelares de las RR Lyrae del tipo c] {\small  Physical parameters for the 
RR1 stars }
\label{fisicosC}
\hspace{0.01cm}
 \begin{tabular}{lccccc}
\hline 
Star& [Fe/H]$_{ZW}$ & $M_V$ & log$(L/L_{\odot})$ & $T_{\rm eff}$ &$D$ (kpc)\\
\hline
V4  &$-1.97$ &0.507& 1.697& 7158.& 18.49\\
V55 &$-1.71$ &0.435& 1.726& 7146.& 17.41\\
V56 &$-1.83$ &0.540& 1.684& 7266.& 18.11\\
\hline
average & -1.84& 0.49 & 1.702 & 7190. &  18.02 \\
$\sigma$ & $\pm$0.13 & $\pm$0.05&$\pm$0.021 & $\pm$66. & $\pm$0.55 \\
\hline
\end{tabular}
\end{center}
}
\end{table}

\subsection{[Fe/H] discrepancies}
\label{sec:discrep}

The internal error of our estimations of [Fe/H] reported in Tables \ref{fisicosAB} and \ref{fisicosC} are small, which
is due to the high quality of the light curves and the Fourier representations, as can also be seen from the small 
uncertainties in Table \ref{tab:fourier_coeffs}. However our [Fe/H] estimate differs from other determinations for
the cluster. The values of [Fe/H] for NGC~5024 reported in the literature have been reviewed recently 
by DK09 who pointed out that most values indicate a metallicity lower than $-1.80$ but values below $-2.0$ are not unusual. DK09 have also estimated the metallicity for twelve giants in NGC~5024 employing a photometric method 
and found a value of $-2.12\pm0.05$. It seems then that the average value of $-1.72$ for the RR0 stars produced 
by eq. \ref{eq:JK96} is too large by 0.2 to 0.3 dex.

The fact that eq. \ref{eq:JK96} predicts higher metallicities by $\sim$~0.2 dex at the low-metallicity end, has been pointed out in the past (e.g. Nemec 2004; Jurcsik \& Kov\'acs 1996; Kov\'acs 2002; Arellano Ferro et al . 2010). A good agreement between the [Fe/H]($P,\phi_{31}$)  and the spectroscopic values for the LMC RR Lyrae stars has been found by Gratton et al. (2004) and Di Fabrizio et al. (2005) only after this metallicity scale difference is taken into account.

Similar systematic offset has been found for the RR Lyraes in metal poor clusters, e.g. Arellano Ferro et al. (2006) for M15;
Nemec (2004) and Arellano Ferro et al. (2008) for NGC 5466; and Arellano Ferro et al. (2010) for NGC~5053. 
This problem has been addressed in detail most recently by DK09, who offer
a series of possible explanations, such as the low number of low-metallicity field RR Lyrae stars in the sample of calibrators
for eq. \ref{eq:JK96} for which no high-dispersion spectroscopy is available. On the other hand the metallicity for the RR1
stars from
eq. \ref{eq:Morgan07}, $-1.84$ seems to predict metallicities closer  independent spectroscopic estimations
(see also the case of M15, NGC~5053 and NGC 5466 in the above cited papers). DK09 have
pointed out that in using eq. \ref{eq:Morgan07} one has to keep in mind that RR1 stars contain less distinct features than RR0 stars; due to the lack of individual spectroscopic
metallicity values for the 106 calibrator stars, the mean values of the parent cluster had to be adopted and as a result only 12 independent metallicity values were considered in the calibration; also the metallicity range 
of the sample is about half of the one covered by the calibration of  eq. \ref{eq:JK96}. DK09 attribute 
the good overall agreement between metallicities to
the strong dependence of [Fe/H] on the period for both RR0 and RR1 stars.

Given the above reasons and the fact that our result for the RR1 stars is based only on three stars and that the
amplitudes of most RR1 stars in NGC~5024 display peculiarities (many Blazhko variables and large scatter in the 
Bailey's diagram) we do not consider the RR1 metallicity as very reliable. On the other hand, 
a systematic correction of 0.2 dex to the Fourier estimate of [Fe/H] for RR0 stars in extreme low metallicity globular clusters leads to a mean value
[Fe/H] = $-1.92 \pm 0.06$.

\subsection{Distance to NGC~5024 from the RR Lyraes stars}
\label{sec:distance}

The  mean absolute magnitudes M$_V$ for the RR0 and the RR1 stars are not significantly different.
The true distance modulus, as calculated for the RR0 and RR1 stars is $16.36\pm 0.05$ and $16.28\pm 0.07$ 
respectively  which correspond to the distances
$18.7 \pm 0.4$ and $18.0\pm 0.5$ kpc, where the quoted uncertainties are the standard deviation of the 
mean. These results
are sensitive to the adopted values of reddening and the ratio $R=A_V/E(B-V)$. We have adopted $E(B-V)=0.02$ 
(Harris 2010, Schlegel et al. 1998) and $R=3.1$ (Clayton \& Cardelli 1988).
Since the above distances for the RR0 and RR1 stars are calculated from two independent empirical 
calibrations, with their own systematic uncertainties, they should be considered as two separate estimates.
The distance to NGC~5024 listed in the catalogue of Harris (1996) is 17.9 kpc, estimated from the mean 
$V$ magnitude of 19 RR Lyraes stars from the work of K00.

In order to provide an independent test to the distance reported above, we have
considered the period-luminosity relation for RR Lyrae stars in the plane $\log~P-I$. 
Fig. \ref{fig:PL_I} shows the distribution of the 
RR Lyrae stars in NGC~5024 on this plane. The colours and symbols are as in Fig. \ref{CMD}. The
RR1 stars periods were fundamentalized using $P_1/P_0 = 0.7454$ (Catelan 2009).
Then we have used the theoretical period-luminosity relation in the $I$ band, calculated by Catelan et al. (2004). 
These authors offer PL relations of the form $M_I = b ~{\rm log}~P + a$ for several filters, values of the metallicity 
and HB type $\cal L$. We used their Table 5 for $Z=0.0005$ to interpolate for $\cal L$$=0.81$ (Rey et al. 1998)
and found the relation
$M_I = -1.385 \log P - 0.301$. Then we have shifted this relation to the observational plane according to the mean distance 
18.7 kpc from the RR0 stars (section 4.5) and
adopting $E(B-V)=0.02$ and $A_I=0.482 A_V$ (Rieke \& Lebofsky 1985). The result is the solid line in Fig. \ref{fig:PL_I}
which matches the observations very well. The data dispersion in Fig. \ref{fig:PL_I}
corresponds to an average uncertainty in the distance of $\pm 0.7$ kpc.

In their paper Catelan et al. (2004) offer average versions of the PL relation for the 
$IJHK$ bands. For the $I$ band they propose; $M_I = 0.471 -1.132 {\rm log~P} + 0.205 {\rm log~Z}$. We calculated 
log~Z from the equation log Z=[Fe/H]zw-1.765+ log (0.638 f+ 0.362) (Catelan et al. 2004) with 
f=$10^{[\alpha/{\rm Fe}]}$=1.995 corresponding to 
[$\alpha$/Fe]=+0.3 (Carney 1966). The resulting relation was shifted to the data
as described in the previous paragraph. 
The result is the segmented line in Fig. \ref{fig:PL_I} which fails matching our data. This illustrates the importance of using the 
HB structure dependent versions of the PL relation whenever the $\cal L$ parameter can be estimated.

\begin{figure} 
\includegraphics[width=7.9cm,height=7.9cm]{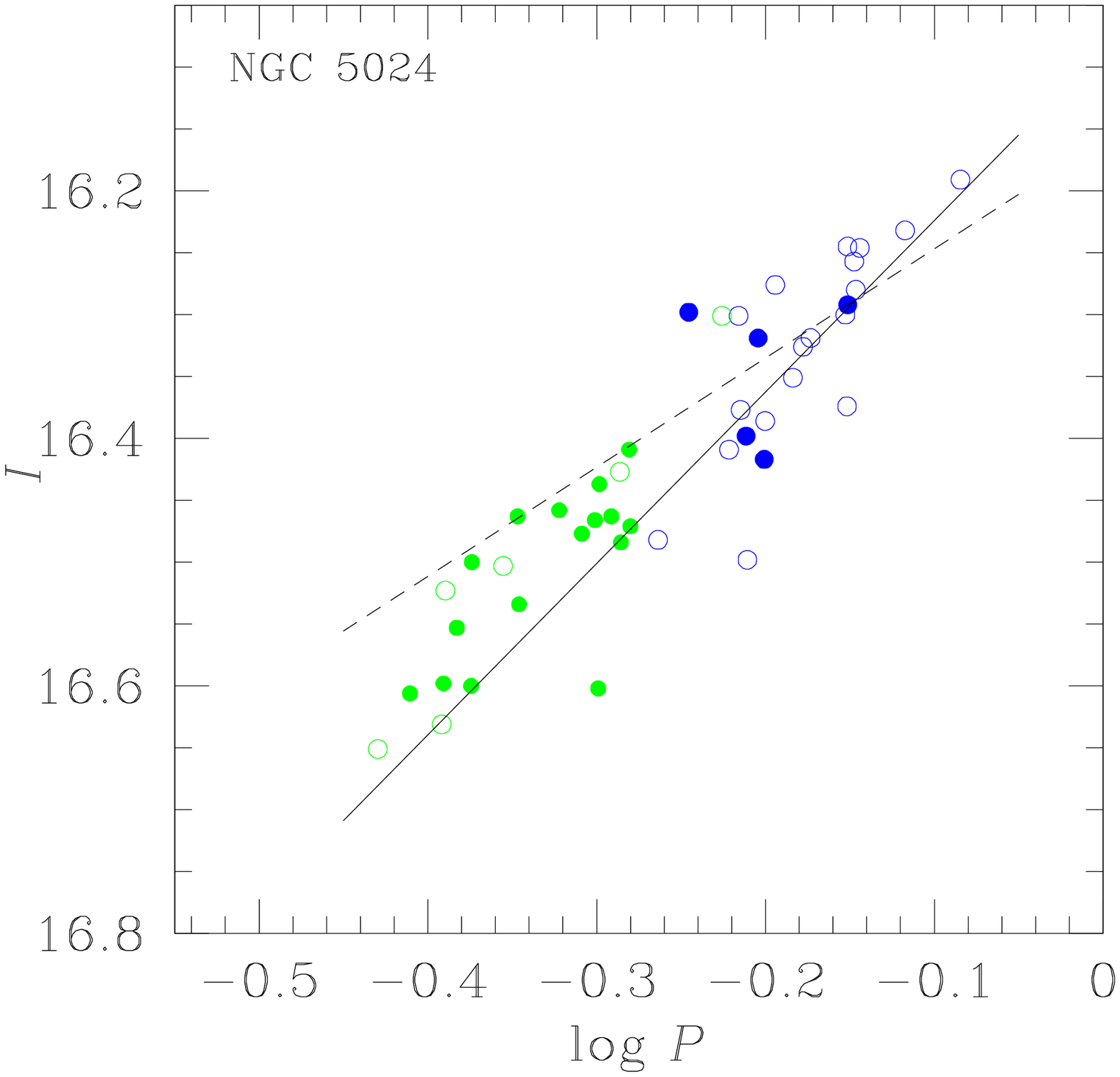}
\caption{Distribution of the RR Lyrae stars in NGC~5024 on the $\log~P-I$ plane. Symbols and colours are as in Fig.\ref{CMD}. Solid line is not a fit to the data but the theoretical PL(I) from Catelan et al. (2004) which has been interpolated for Z=0.0005 and $\cal L$$=0.81$, after shifting it to the distance and reddening of the cluster. The dashed line correspond to the average PL(I) relation $M_I = 0.471 -1.132 {\rm log~P} + 0.205 \rm{log~Z}$ of Catelan et al. (2004) equally shfted to the data.
See section 4.5 for a discussion.}
    \label{fig:PL_I}
\end{figure}

\subsection{Masses for the RR0 stars}

The masses of the RR0 stars with stable light curves can be estimated from the calibration of Jurcsik (1998);

$log~M/M_{\odot} = 20.884 - 1.754~\log P + 1.477~\log~(L/L_{\odot})$
\begin{equation}
- 6.272~\log~T_{\rm eff} + 0.0367~{\rm [Fe/H]}.
\label{eq:mass}
\end{equation}

Although the validity of this equation has been questioned by Cacciari et al. (2005), in a recent paper (Arellano Ferro et al. 2010) we have compared the masses from eq. \ref{eq:mass}
with the predicted masses from the van Albada \& Baker (1971):  $\log~M/M_{\odot} = 16.907 - 1.47~ \log~P_F + 1.24~\log~(L/L_{\odot}) - 5.12~\log~T_{\rm eff}$ for a sample of RR0 stars in NGC~5053, and found that they agree to within a few percent. We have calculated the masses of the RR0 stars in NGC~5024 with eq. \ref{eq:mass} and report them in the last column of Table 
\ref{fisicosAB}. The average mass is $0.72\pm 0.02~M_{\odot}$

We did not calculate the masses for RR1 stars due to the amplitude and phase modulations  discussed in previous sections for these stars.

\begin{figure*} 
\includegraphics[scale=1.0]{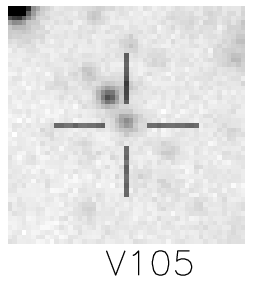}
\caption{Detailed identification of the 13 newly found SX Phe stars. Their positions in the general field of the cluster are shown in Fig. \ref{N5024a}.}
    \label{NEWSX}
\end{figure*}

\section{SX Phe stars}
\label{sec:SXPHE}

We have carried out an inspection of the light curves in our data of the four suspected SX Phe by DK09: 
V80, V81, V82 and V83 but have only confirmed the variability and periodicities reported by DK09 for 
V80. V81 and V82 are not found to be variable. The variability of V83 is marginal but its SX Phe nature 
cannot be ascertained. 

\setcounter{figure}{9}
\begin{figure*} 
\includegraphics[width=18.cm,height=22.cm]{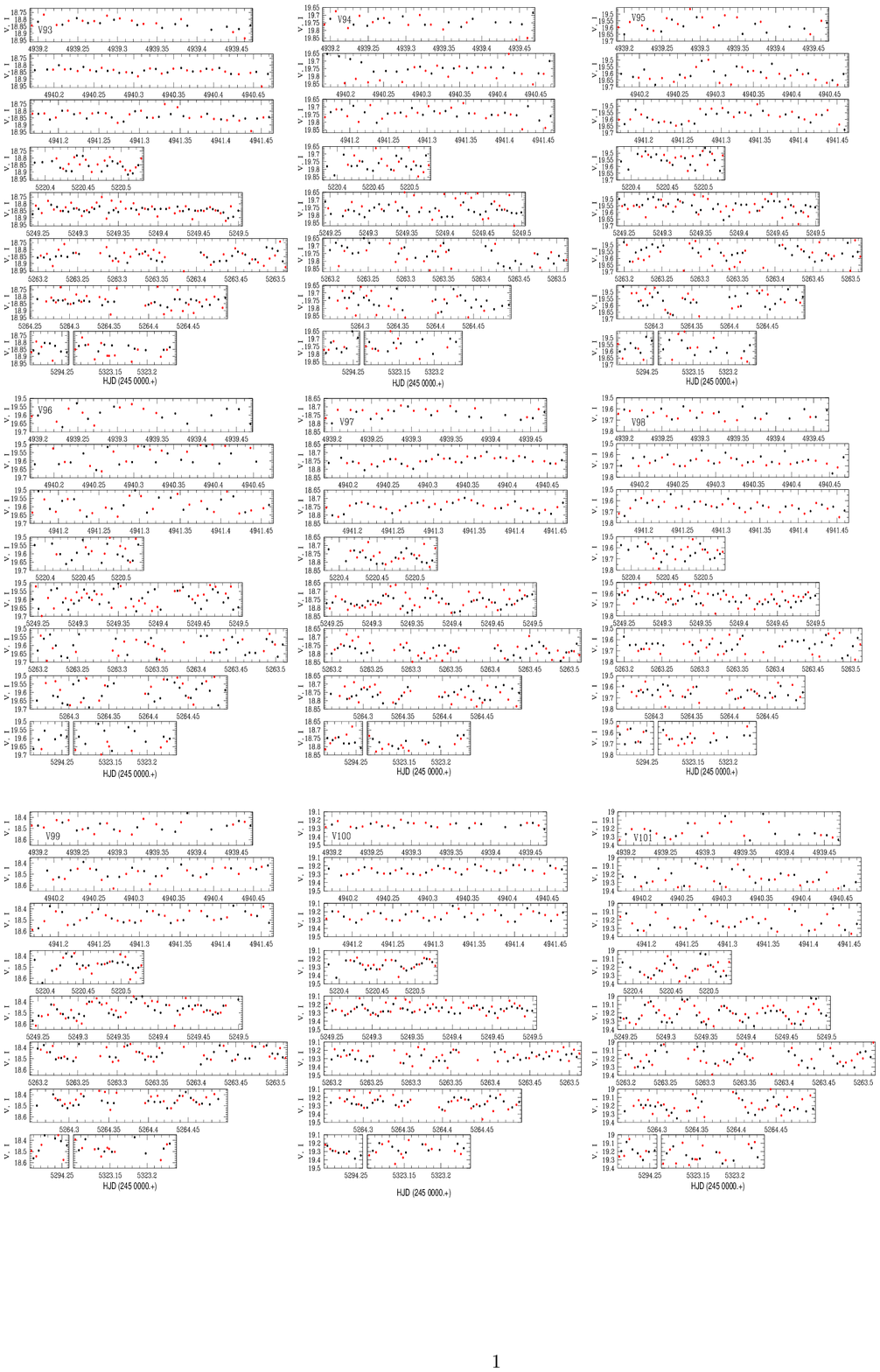}
\caption{Light curves of newly found SX Phe stars.
To produce a more complete appearance of the light curve we have applied an arbitrary shift in magnitude to the $I$ data (red circles) to match the $V$ data (black circles).}
    \label{fig:SX_LC}
\end{figure*}

\setcounter{figure}{9}
\begin{figure*} 
\includegraphics[width=18.cm,height=14.cm]{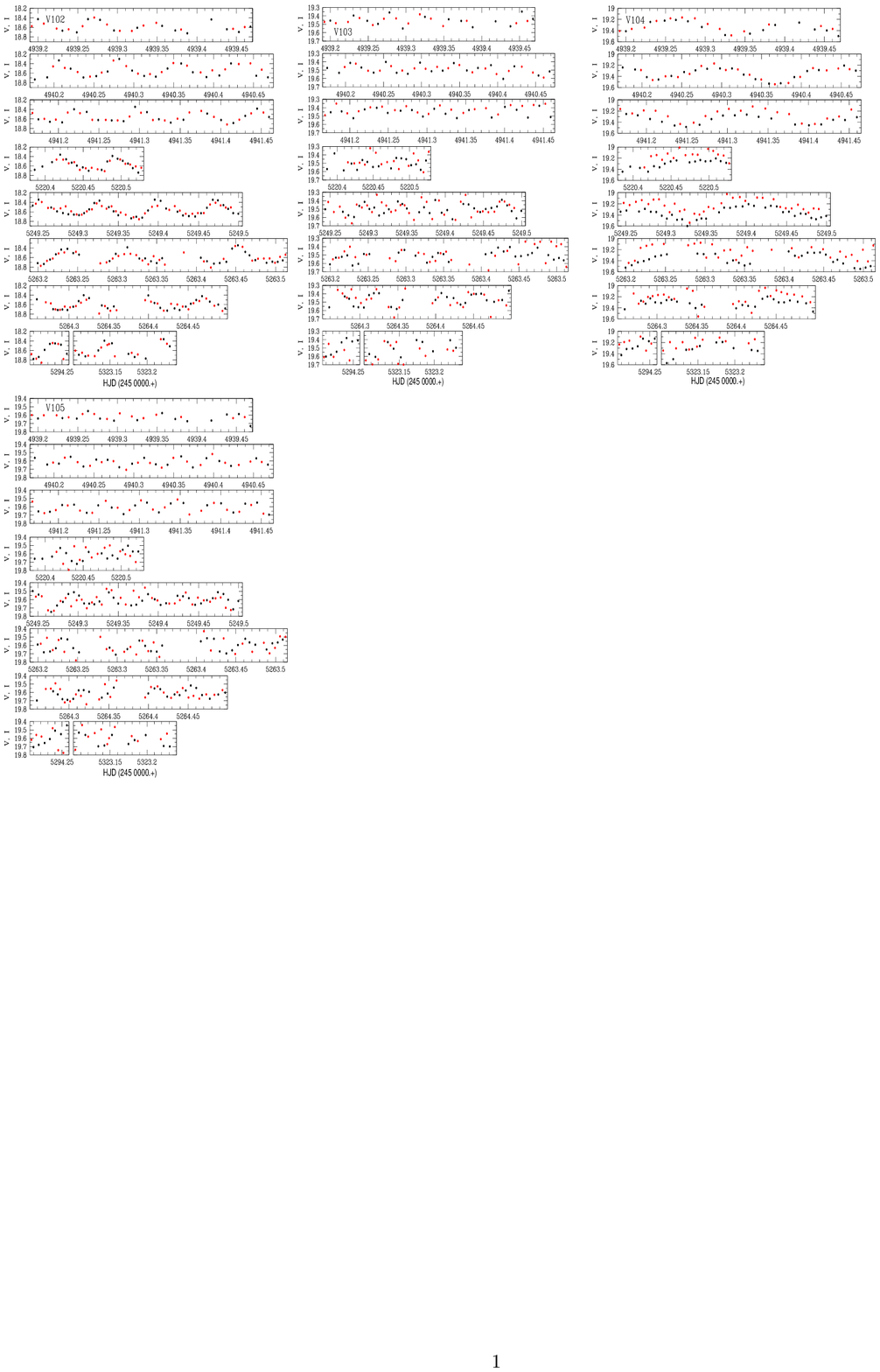}
\caption{Continued}
\end{figure*}

In the present work 13 new SX Phe stars have been found and another 5 marginal cases are retained as suspected. 
The light curves of some of these new SX Phe stars are shown in Fig \ref{fig:SX_LC}. Accurate finding charts for the
these new SX Phe stars are given in Fig. \ref{NEWSX}. We have performed a 
period analysis for all these stars with {\tt period04}. Due to the limited time span of our data, 
it is not possible to assess the existence 
and values for more than one, and in some cases two, frequencies.
However the determination of the main frequency with the largest amplitude is quite clear. 
In Fig \ref{newSXPHE} we present the frequency spectra for all the newly found SX Phe stars.

The corresponding variable names assigned to the new variables are from V93 to V105. 
Their frequencies, mode identification and celestial coordinates are given in Table \ref{tab:SXPHE}. 
For the mode identification we have assumed that the main frequency, with the largest amplitude, is
the fundamental mode $f_1$. When secondary frequencies are identified, the mode is suggested by the
frequency ratio (e.g. V99, V101, V102). However the final mode assignment has been defined by the
position of each star on the PL plane of Fig. \ref{BS_PL} relative to the fundamental and first overtone
loci as discussed the following section.

\begin{table*}
\caption{Frequencies and amplitudes of the newly found SX Phe stars. The numbers in parentheses indicate the uncertainty on the last decimal places.}
\centering
\begin{tabular}{lccrccccl}
\hline

Variable & $V$& $V-I$ & \multicolumn{1}{c}{$f$}& $A$ &$P$ &RA & Dec. & mode id.\\
          & & & \multicolumn{1}{c}{(d$^{-1}$)} &  (mag) & (d)& J2000.0  & J2000.0&\\
\hline
V93$^a$&18.840&0.303&24.95330(17)&0.014(2)&0.04007 & 13 12 40.00 & +18 08 39.0&$f_1; 2O$ \\
V94 &19.750 &0.333 &26.32849(21)&0.022(4)&0.03798 &13 12 42.11  & +18 07 39.8 &$f_1; F$ \\
V95 &19.564 &0.290 &24.36281(18)&0.029(5)&0.04105 &13 12 47.94 &+18 09 55.2 &$f_1; F$\\
    &       &      &13.95330(30)&0.017(5)&0.07167 & &  & $f_2; non-radial$\\
V96 &19.589 &0.213 &25.46587(6)&0.063(3)&0.03927 &13 12 48.29 &+18 13 18.9&$f_1; F $\\
V97 &18.762 &0.309 &16.61477(3)&0.033(2)&0.06019 &13 12 48.27 &+18 14 34.6&$f_1; 1O$\\
V98 &19.642 &0.351 &26.04022(11)&0.040(4)&0.03840 &13 12 48.67 &+18 09 35.9&$f_1; F$\\
V99 &18.456 &0.019 &17.72343(6)&0.064(4)&0.05642 &13 12 52.91 &+18 10 35.8&$f_1; F$\\
    &       &      &22.66610(15)&0.028(4)&0.04412 & & &$f_2; 1O$; $f_1/f_2 = 0.782$ \\
V100&19.264 &0.295 &20.74930(8)&0.045(3)&0.04819 &13 12 53.42 &+18 14 46.2&$f_1; F$\\
    &       &      &21.00011(22)&0.016(4)&0.04762 & & &$f_2;  non-radial$ \\
V101&19.211 &0.179 &19.04200(4)&0.096(4)&0.05252 &13 12 53.65 &+18 08 57.9&$f_; F$\\
    &        &     &24.34878(7)&0.061(4)&0.04107 & & &$f_2; 1O$; $f_1/f_2 = 0.782$ \\
V102&18.576  &0.318&13.80387(3)&0.153(4)&0.07244 &13 12 54.76 &+18 09 37.7&$f_1; 1O?$ \\
     &       &     &27.60780(9)&0.047(4)&0.03622 & & &$f_2$; $2f_1$ \\
     &       &     &21.18330(24)&0.016(4)&0.04721 & & &$f_3: 2O?$; $f_1/f_3 = 0.652$ \\
V103 &19.484 &0.306&22.56317(8)&0.068(5)&0.04432 &13 12 58.32 &+18 08 41.2 &$f_1; F$ \\ 
V104 &19.342 &0.174& 6.77748(30)&0.115(10)&0.14756 &13 12 59.00 &+18 10 22.0 &$f_1; 3f_1= F?$\\
V105 &19.609 &0.310&22.09475(6)&0.067(4)&0.04526 &13 13 01.92  &+18 12 30.4 &$f_1; F$ \\

\hline
\end{tabular}
\label{tab:SXPHE}
\\ \flushleft$^a$although clear variations are seen and a distinctive peak is seen in the frequency spectrum, the amplitude is very small.\\
$F$:Fundamental mode; $1O$: First overtone; $2O$: Second overtone.
\end{table*}

\subsection{SX Phe PL relationship}
    \label{sec:SXPHE_PL}

The existence of a period-luminosity relationship for SX Phe stars has been firmly established both 
from theoretical (e.g. Santolamazza et al. 2001) and empirical grounds (e.g. McNamara 1997; 2000, 
Jeon et al. 2003; 2004, Poretti et al. 2008). In Fig. \ref{BS_PL} we have plotted on the 
log~$P -V$ and log~$P -I$ planes, the previously known and suspected SX Phe stars, as given 
in the CVSGC, and the 13 SX Phe stars stars found in this work. The solid lines in both panels are least square fits 
to the evident fundamental pulsator stars. Small and large dashed lines correspond to relationships 
inferred for first and second overtone pulsators respectively,
assuming the ratios $F/1O = 0.783$ and $F/2O = 0.571$ (see Santolamazza et al. 2001 or Jeon et al. 2003; 
Poretti et al. 2005).
The distribution of the stars along one of these PL relations is very clear in most cases. 
On the basis of this distribution and the frequency ratios from the period analysis we have assigned 
the pulsation mode given in column 9 of Table  
\ref{tab:SXPHE}.  Although the stars V78 and V90 are not in the field of our images, we included 
them by adopting their periods and $V$ magnitudes from the CVSGC. The previously suspected SX Phe 
star V81 is not included since we have not confirmed its variability. Two positions are plotted for 
V101 and V104.
V101 is a clear case for which the fundamental mode and first overtone have been detected. For V104 
however, the main frequency found is too short for its magnitude both in $V$ and $I$. If we assume that we have detected 
an alias and that the fundamental frequency $F= 3f_1$, then the star positions right on the fundamental
sequence of the PL relationship. We concede that this is forcing the case but it is tempting to 
assume that. In any case we have not used V104 in the calculation of the fundamental 
log~$P -V$ relation.
There are outstanding outliers such as V78, and the two of the suspected SX Phe stars
V82 and V83. It is noted that these three stars fall outside the BS region on the CMD of 
Fig. \ref{CMD}. The variability of
V78 is clear but it is too bright for its period. We speculate that the star has an unseen companion 
that makes it appear brighter. The star V83 in filter $I$ falls sharp on the
fundamental sequence. The case of V82 and V83 has been addressed in section \ref{sec:IDVAR} 
and their SX Phe nature has not been confirmed. 

Now, the above procedure has produced some peculiarities. For instance, for the star V99 the frequency 
analysis showed very clearly two frequencies whose ratio $f_1/f_2=0.782$ is the expected one between 
the fundamental and first overtone for metallicities [Fe/H]$< -1.0$ (e.g. Fig. 4 in Poretti et al. 2005). 
However, when plotted on the log~$P -V$ plane of Fig. \ref{BS_PL} with the 
principal period, the star falls on the second overtone locus and on the log~$P -I$ 
plane on the first overtone! It is likely that the star is a blend. A similar case is
that of V102. Both stars appear to be too bright for their putative fundamental period. Although we do 
not have a ready explanation for this, it may
be possible that the stars are not SX Phe and hence do not follow the corresponding PL relation 
and/or that they are 
members of a binary system and their magnitude is contaminated by the companion. 

\begin{figure} 
\includegraphics[width=8.cm,height=16.cm]{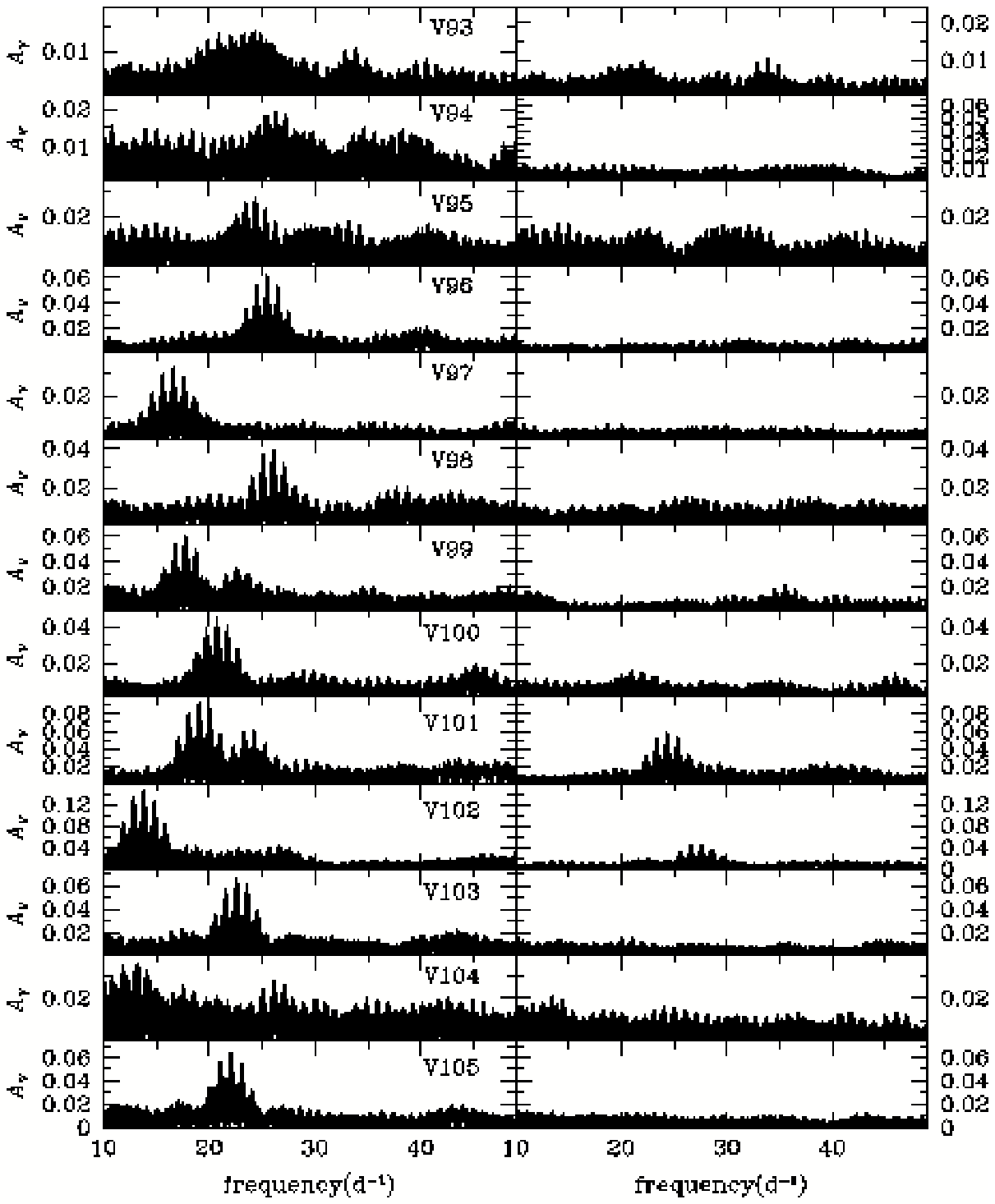}
\caption{Frequency spectra of 13 newly identified SX Phe stars. The spectra on the left
are calculated from the original data. The spectra on the right have been prewhitened from the major frequency. In some cases sufficient signal remains to identify secondary frequencies. See Table \ref{tab:SXPHE} for the frequencies and amplitude values.}
    \label{newSXPHE}
\end{figure}

\begin{figure}  
\includegraphics[width=8.cm,height=12.5cm]{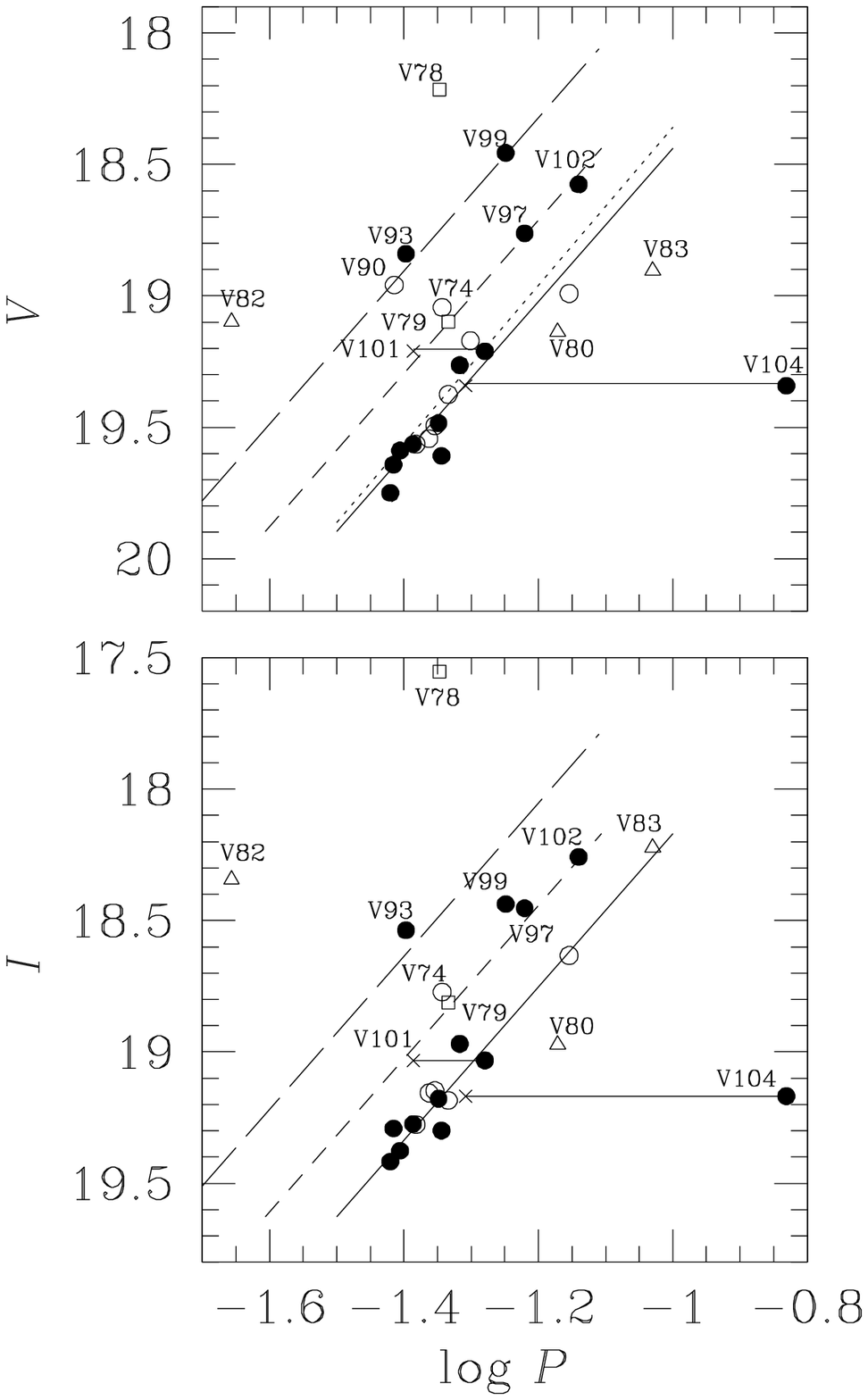}
\caption{Period-Luminosity relationship in $V$ and $I$ for SX Phe stars in NGC 5024. Empty circles represent the
8 SX Phe discovered by Jeon et al. (2003). Filled circles are the 13 SX Phe stars discovered in this work. Empty squares are two SX Phe stars found by DK09. 
Triangles are three of the four stars suspected by DK09 to be SX Phe star. V81 is not included since we 
believe it is not a variable star.
The solid lines is a least square fit to the evident fundamental pulsator stars. Short and long dashed 
lines correspond to the relationships inferred for first and second overtone pulsators respectively,
assuming the ratios $F/1O = 0.783$ and $F/2O = 0.571$. See section \ref{sec:SXPHE_PL} for a detailed discussion. The dotted line in the top panel is the P-L calibration calculated by Jeon et al. (2003) from 6 SX Phe in NGC~5024.}
    \label{BS_PL}
\end{figure}

All the peculiar stars were ignored in the calculation of the least squares fits of the apparently fundamentally pulsating stars, the solid lines in Fig. \ref{BS_PL}.
The fundamental log~$P - V $ and log~$P - I $ relationships are of the form:

\begin{equation}
V = (-2.916\pm 0.350)~  \log P + (15.524 \pm 0.511),
\label{eqn:SX_PL}
\end{equation}

\begin{equation}
I= (-2.892\pm 0.188)~  \log P + (15.318 \pm 0.269),
\label{eqn:SX_PL}
\end{equation}

\noindent which, by adopting the average distance modulus obtained from the RR0  stars 
$\mu_o$ = $16.36 \pm 0.05$ and $E(B-V)=0.02$ (section \ref{sec:distance}) correspond to:  

\begin{equation}
M_V = -2.916~ \log P - 0.898. 
\label{eqn:SX_PMv}
\end{equation}

\begin{equation}
M_I = -2.892~ \log P - 1.072. 
\label{eqn:SX_PMi}
\end{equation}

Eq. \ref{eqn:SX_PMv} can be compared with the one found by Jeon et al. (2003) for NGC~5024 based 
only on the six fundamental SX Phe stars then known: $M_V= -3.010~{\log P} - 1.070$. This 
relationship is also shown in 
Fig. \ref{BS_PL} as a dotted line. The new $M_V$ calibration has essentially the same slope but it is 
fainter by about 0.05 mag.

The slope and zero point of the PL relation for SX Phe have been obtained for several clusters
and discussed by Jeon et al. (2003)
(see for instance their Fig. 7). The slope for M55 ($-2.88\pm0.17$ Pych et al. 2001) and for NGC 5024  ($-3.01\pm0.262$ Jeon et al. (2003)) and also the slopes from theoretical studies ($-3.05$ Santolamazza et al. 2001; $-3.14$ Templeton et al. 2002) are, within the uncertainties, similar to our value in eq. \ref{eqn:SX_PMv}. 
Slightly larger values have been obtained by recent analysis of 153 Galactic and extragalactic stars $\delta$ Scuti and SX Phe stars, $-3.65\pm0.07$ (Poretti et al. 2008), and $-3.725\pm0.089$ for 29 Galactic $\delta$ Scuti and SX Phe stars (McNamara 1997). It is intriguing that
the Galactic globular cluster with the largest number of known SX Phe stars, $\omega$ Cen (Kaluzny et al. 1997), displays a slope as steep as $-4.66$ (McNamara 2000).
It is likely that a precise PL relation conveys a metallicity dependence, hence studying homogeneous samples in globular clusters is of great relevance. Unfortunately the population of known 
SX Phe in globular clusters is generally small. The discovery of 13 new SX Phe in NGC~5024
nearly triples the number and allows to firmly include this cluster in the general discussion 
of the PL relation calibration. It is comforting to see that the resulting PL relation found above is in a good agreement with independent previous results.

\subsection{Suspected Variables}

A few stars whose light curves show mild but rather clear variations were identified. 
However, we failed to find
a convincing frequency above the noise level. These stars are retained as suspected variables and they have been labeled with an "S" prefix. They are listed in Table \ref{suspected} along with their celestial coordinates. One should refrain from assigning these stars a variable number until their variability can be confirmed.

\begin{table}
\caption{Blue stragglers whose variability is to be confirmed.}
\centering
\begin{tabular}{llcll}
\hline
Star & $V$&$V-I$ &  RA        & Dec.   \\
     & & & J(2000.0)  & J(2000.0)\\
\hline
S1 &19.977 &0.055 &13 12 47.0 &+18 10 22.8 \\
S2 &18.998 &0.320 &13 12 49.3 &+18 10 05.6 \\
S3 &18.163 &$-0.104$ &13 12 51.0 &+18 10 04.3 \\
S4 &19.866 &0.31  &13 12 53.9 &+18 09 15.1 \\
S5 &18.084 &-0.316 &13 12 56.0 &+18 09 14.4 \\
\hline
\end{tabular}
\label{suspected}
\end{table}

\section{Long Term Variables}
\label{sec:LongTerm}

Of the known long term variables in NGC~5024, only 
three are not saturated; V67, V84, and V86. V84
falls very close to the edge of our field, and a CCD defect 
has randomly affected the images of this star. Hence we do not claim good observations for V84. 
For both V67 and V86 our $V$ and $I$ 
photometry is reliable keeping in mind that such bright stars usually suffer from systematic 
problems with the photometry to a greater extend than most other stars.

We have attempted period determinations for these stars using {\tt period04}. 
For V67 and V86 no $V$ photometry is given by DK09. 
The analysis of our data suggests periods of 29.4 days for V67 and 12.2 days for V86. 
No other period is known for V67 while for V86
DK09 report a period of 22.2 days. We note that neither 12.2 nor 22.2 days produce a 
reliable folding of our data but also note that the time span or our data set is much 
shorter than that of DK09, thus it is 
likely that their period is closer to the reality.

V85 is saturated in the filter $I$ reference image and hence we publish only 
reliable $V$ data, 
that have been used to discuss its periodicity below. $V$ data are available from DK09 and we
have noticed a small shift of about 0.04 mag between the
two data sets. Since applying a correction would be inaccurate and would affect the period
analysis, we chose to analyze the two data sets separately. We have confirmed the 19.8 day period reported
by DK09 based on their data set alone and found that this period phases well our data.

In Table \ref{tab:LPV} we report the mean $V$ magnitude and colour 
$V-I$ which in fact place the stars 
correctly on the RGB of the CMD as shown in Fig. \ref{CMD}.

\begin{table}
\caption{Mean $V,I$ photometry of long period variables}
\centering
\begin{tabular}{llc}
\hline
Star & $V$&$V-I$ \\

\hline
V67 &14.221 &1.295  \\
V84 &14.720 &1.176  \\
V85 &14.024 &--     \\
V86 &14.154 &1.352  \\
\hline
\end{tabular}
\label{tab:LPV}
\end{table}

\section{On the age of NGC~5024 and a reddening estimation}
\label{sec:age}

Ages of globular clusters have been estimated from the so called "vertical" and "horizontal" 
methods, i.e. estimating $\Delta$$V$, the difference of magnitudes of the ZAHB and the Turn 
Off (TO) point and $\Delta$$(V-I)$ or $\Delta$$(B-V)$, the difference between the TO and a 
fiducial point on the RGB (Rosenberg et al. 1999; Salaris \& Weiss 2002).  
Both $\Delta$$V$ and $\Delta$$(V-I)$ are age and metallicity dependent. The usage of any of these 
methods to the CMD of NGC~5024 in Fig. \ref{CMD} faces the 
immediate problem of locating the TO point given the large scatter at $V \sim 20$ mag. 
Therefore,
instead of attempting an independent determination of the age of the cluster, we have adopted the recent
age estimate of 13.25$\pm$0.50 Gyr reported by Dotter et al. (2010) which was achieved from isochrone 
fitting on a deep CMD based on Hubble Space Telescope, Advanced Camera Survey photometry.

We have plotted on the CMD of Fig. \ref{CMD} the isochrones from the library of
VandenBerg, Bergbusch \& Dowler (2006) for [Fe/H]=$-1.84$ (purple lines) and $-2.01$
(green lines), [$\alpha$/Fe]=$+0.3$
and the ages 12 (continuous lines) and 14 Gyr (segmented lines), which bracket the calculated metallicity from the RR0 stars 
and the adopted age for the cluster. The isochrones were shifted 
until a satisfactory fit to the RGB was observed. We found the best solution 
for an apparent modulus $\mu$=16.42, and $E(B-V)=0.0$ which are consistent
with our estimated true distance modulus $\mu_0 = 16.36$ and the adopted $E(B-V)=0.02$ in section 4.5. The CMD in Fig. \ref{CMD} is consistent the [Fe/H] and distance values
obtained from the Fourier decomposition approach to the RR0 stars, and the age of 13.25$\pm$0.50 Gyr adopted from Dotter et al. (2010).

\section{Conclusions}

The technique of difference imaging has proved to be a powerful tool to find new variables even in crowded image regions. Here we report the discovery of two RR1 and 13 SX Phe stars.
These discoveries were possible thanks to the spatial resolution of our images and to the deepness and quality of our time series photometry. The difference imaging approach employed for the image analysis played a major role in improving
the quality of the light curves and the discovery of very faint variables.
This also permitted us to correct previous misidentifications of some variables in the cluster.

Physical parameters of astrophysical relevance, [Fe/H], $M_V$, $\log~(L/L_{\odot})$, 
 $\log~T_{\rm eff}$, and mass, have been derived for selected RR Lyrae stars in NGC~5024
using the Fourier decomposition of their light curves. Special attention has been paid to the 
conversion of these parameters onto broadly accepted scales.

Mean values of [Fe/H] and distance of NGC 5024 were calculated from individual values  of carefully 
selected RR Lyrae stars whose light curves are stable and have been shown to have a good quality. The 
final estimates are [Fe/H]=$-1.92 \pm 0.06$ from 19 RR0 stars and the distance moduli 
$16.36 \pm 0.05$ and $16.28 \pm 0.07$ obtained independently for RR0 and RR1 stars  
respectively, and which correspond to the distances $18.7 \pm 0.4$ and $18.05 \pm 0.5$ kpc. 

We found 13 new SX Phe stars of $V$ magnitudes between 18.4 and 19.7 and their pulsation periods were 
calculated. In a few cases more than one period was detected. The 13 newly discovered SX Phe stars together with the 8 previously known, allowed new calibrations
of the PL relationship of SX Phe stars in the $V$ and $I$ bands, and to assign a pulsation mode for most of these elusive faint 
stars. The calibration of the PL relationship for cluster SX Phe provides a powerful tool for distance 
determination and it agrees with independent calculations in different stellar systems.

\section*{Acknowledgments}

We are grateful to the support astronomers of IAO, at Hanle and CREST (Hosakote) for their very efficient help while acquiring the data. We thankfully acknowledge the numerous comments and suggestions from an anonymous referee. AAF acknowledges the hospitality of the Indian Institute of Astrophysics during his sabbatical leave in 2010. This project was supported
by DGAPA-UNAM grant through project IN114309 and by the INDO-MEXICAN collaborative program by DST-CONACyT. This work has made a large use of the SIMBAD and ADS services.

\section*{Supporting Information}
\label{sec:support_info}

Additional supporting information may be found in the online version of this article.

{\bf Table 1.} Time-series $V$ and $I$ photometry for all the confirmed variables in our field of view.

Two tar files containing the individual  $V$ and $I$ light curves for all stars measured
in the field of our images are also included as online material.

Please note: Wiley-Blackwell are not responsible for the content or functionality of any supporting materials supplied by the authors.
Any queries (other than missing material) should be directed to the corresponding author for the article.

\label{lastpage}

\end{document}